\documentclass[prb,amsfonts,showpacs,preprint]{revtex4}
\usepackage{hyperref,graphics,bm}
\begin{document}
\widetext
\title{Ferromagnetic zigzag chains and properties of the
charge ordered perovskite manganites}
\author{I. V. Solovyev}
\affiliation{
JRCAT-Angstrom Technology Partnership,
1-1-4 Higashi, Tsukuba, Ibaraki 305-0046, Japan\\
and\\
Institute of Metal Physics, Russian Academy of Sciences, \\
Ekaterinburg GSP-170, Russia}
\date{\today}

\widetext
\begin{abstract}
The low-temperature properties of the so-called ''charge ordered'' state in
50\% doped perovskite manganites
are described from the viewpoint of the magnetic spin
ordering. In these systems, the zigzag antiferromagnetic ordering, combined
with the double-exchange physics, effectively divides the whole sample into the
one-dimensional ferromagnetic zigzag chains and results in the anisotropy of
electronic properties. The electronic structure of one such chain is described
by an effective $3$$\times$$3$ Hamiltonian in the basis of Mn($3de_g$) orbitals.
We treat this problem analytically and consider the following properties:
(i) the nearest-neighbor magnetic interactions; (ii) the distribution of the
Mn($3de_g$) and Mn($4p$) states near the Fermi level, and their contribution
to the optical conductivity and the resonant x-ray scattering near the Mn
$K$-absorption edge. We argue that the anisotropy of magnetic interactions in
the double-exchange limit, combined with the isotropic superexchange interactions,
readily explains both the local and the global stability of the zigzag
antiferromagnetic state. The two-fold degeneracy of $e_g$ levels plays a
very important role in the problem and explains the insulating behavior of the
zigzag chain, as well as the appearance of the orbital ordering in the
double-exchange model. Importantly, however, the charge ordering itself
is expected to play only a minor role and is incompatible with the
ferromagnetic coupling within the chain. We also discuss possible effects of
the Jahn-Teller distortion and compare the tight-binding picture with results
of band structure calculations in the local-spin-density approximation.
\end{abstract}

\pacs{PACS numbers: 75.25.+z, 71.45.Lr, 78.20.Bh, 75.30.Vn}

\maketitle

\section{Introduction}
\label{sec:intr}

\begin{figure}
\centering \noindent
\resizebox{10cm}{!}{\includegraphics{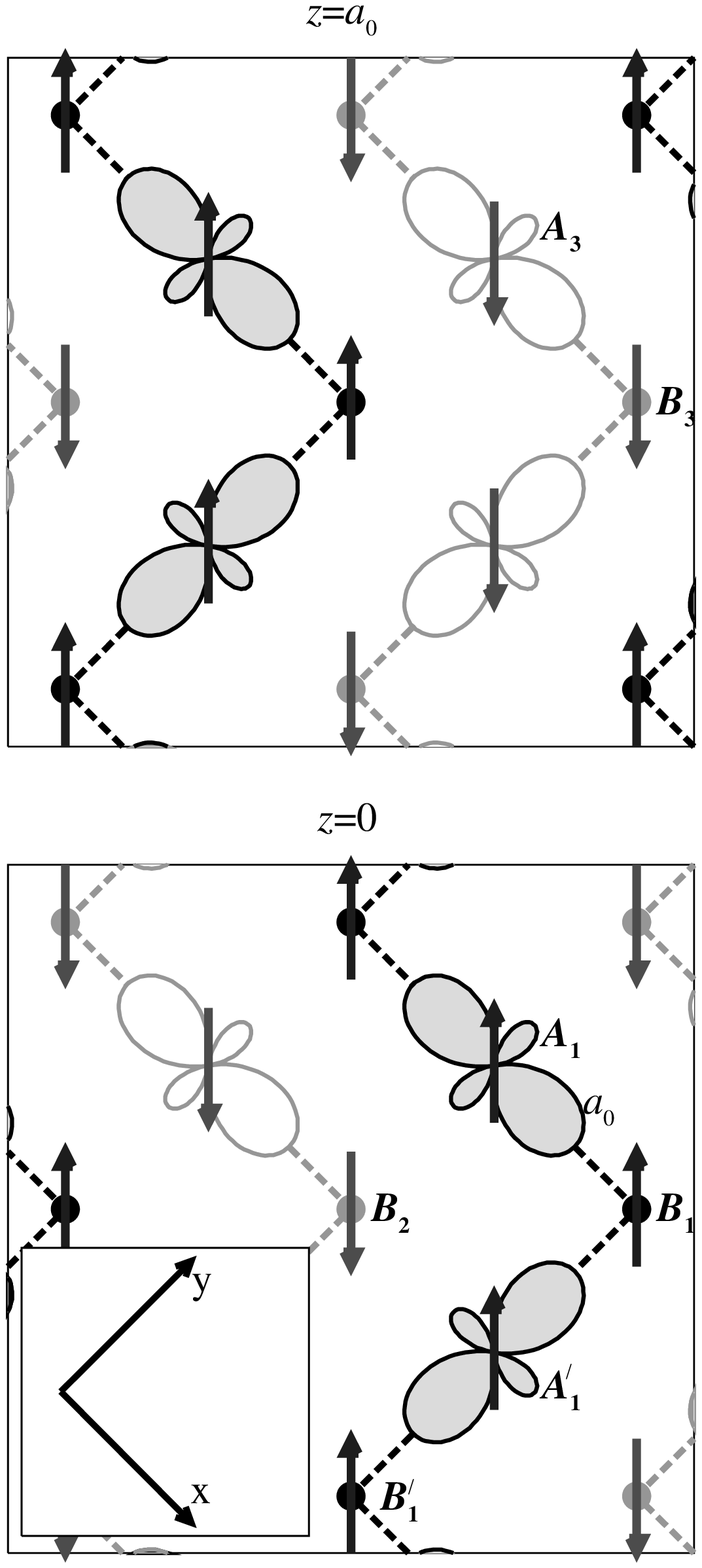}}
\caption{Spin, orbital and ''charge'' ordering in two neighboring planes of
         the CE-type antiferromagnetic structure: symbols $A$ and $A'$
         stand for the bridge sites, symbols $B$ and $B'$ stand
         for the corner sites.
         $a_0$$\equiv$$|A_1B_1|$ is the
         cubic lattice parameter.}
\label{fig.chains}
\end{figure}

  It appears that many perovskite manganites, when doped by holes up to the level
corresponding to the formal valence of Mn $+$$3.5$, exhibit very specific
properties, inherent to only this particular hole concentration and different from
other doping regimes. The compounds are insulating and at certain temperature range
(typically at low temperatures) form rather peculiar magnetic ordering consisting
of ferromagnetic (FM) zigzag chains, coupled antiferromagnetically --
see Fig. \ref{fig.chains}.
In the case of three-dimensional manganites, this is the well known
CE-type antiferromagnetic (AFM) ordering, which
was proposed long ago by Goodenough,\cite{Goodenough}
and first observed by Wollan and Koehler in
La$_{1-x}$Ca$_x$MnO$_3$ near $x$$=$$0.5$.\cite{WollanKoehler}
The zigzag AFM ordering appeared to be a more generic
and was observed recently in a number of systems,
including the layered manganite La$_{1/2}$Sr$_{3/2}$MnO$_4$
(Refs. \onlinecite{Sternlieb,Murakami1})
and the double-layer manganite LaSr$_2$Mn$_2$O$_7$ (Ref. \onlinecite{Kubota}).
The magnetic ordering is accompanied by the orbital ordering,\cite{Murakami1}
which is also shown
in Fig. \ref{fig.chains}. The insulating zigzag AFM state can be
transformed into the metallic FM state by applying an external magnetic
field.\cite{melting} Other examples of such compounds and detailed description of
their properties can be found in the review articles.\cite{review.1}

  The systems are usually referred to as the ''charge ordered'' manganites, implying
that the homogeneous state of Mn$^{3.5+}$ ions in the lattice is unstable towards the
charge disproportionation amongst two alternately aligned types of sites:
$2$Mn$^{3.5+}$$\rightarrow$Mn$^{4+}$$+$Mn$^{3+}$. The electronic configurations of
the Mn$^{4+}$ and Mn$^{3+}$ ions are $t_{2g}^3$ and $t_{2g}^3e_g^1$, respectively.
Thus, the $t_{2g}$ orbitals are half-filled in both cases, giving rise to the
AFM superexchange interaction between nearest neighbors. Due to the
strong
Hund's rule coubling, the $e_g$ electron and three $t_{2g}$ electrons
at the Mn$^{3+}$ site are aligned ferromagnetically. It is assumed that the
$e_g$ electron is mobile and can hop to the neighboring Mn$^{4+}$ by forming the
FM double-exchange (DE) coupling in a certain number of Mn$^{3+}$-Mn$^{4+}$
bonds.\cite{Anderson} The occupied $e_g$ orbitals are ordered in the direction of
these bonds so to maximize the DE coupling.
This is the basic physical idea underlying Goodenough's conjecture for the CE-type
AFM ordering.\cite{Goodenough}
The picture had no solid justification and
for many years remained to be on the level
of a hypothesis.

  Only very recently it was realized by several authors
that, at least at very low
temperatures, the basic proporties of the ''charge ordered'' manganites
can be understood exclusively from the viewpoint
of the DE physics.\cite{Hotta,Solovyev1,Brink}
Due to the strong Hund's
rule coupling, the Mn ions form a high-spin state in manganites.
Therefore, the intraatomic exchange splitting is large.
If it is much larger than the kinetic hoppings
between nearest-neighbor $e_g$ orbitals, the minority-spin $e_g$ states can be
projected out of the problem. In the case of AFM ordering, this
will suppress all kinetic hoppings between Mn-sites with
opposite directions of the spin magnetic moments. In the following we will refer to
this picture as the DE limit. Thus, the CE phase in the DE
limit will be effectively devided into the one-dimensional zigzag chains.
This will create two geometrically different types of Mn-sites: the
so-called bridge sites $A$ and $A'$, and the corner sites $B$ and $B'$
-- see Fig. \ref{fig.chains}.
However, the difference is
not necessarily
related with the integer change of atomic charges when going from one type of
Mn-sites
to another.
In the other words, the bridge sites are not necessarily occupied
by the Mn$^{3+}$ ions and the corner sites - by the Mn$^{4+}$ ions.
The two-fold degeneracy of $e_g$ levels plays a very important role in the
problem.
The isolated chain
behaves as a band insulator.
The only occupied $e_g$ orbitals at the sites $A$ and $A'$
are $3x^2$$-$$r^2$ and $3y^2$$-$$r^2$, respectively, that is exactly the
orbital ordering shown in Fig. \ref{fig.chains}.
In our recent Letter (Ref. \onlinecite{Solovyev1}) we argued that the local stability
of the zigzag AFM state can be explained by the anisotropy of the interatomic
magnetic interactions in the DE limit, whereas the charge ordering
itself
may be irrelevant to the problem.

  In the present paper we will further ellaborate this DE picture.
In Sec. \ref{sec:tbmodel} we will derive the tight-binding (TB) Hamiltonian for the
zigzag chain, discuss basic features of the electronic structure, and
obtain several useful expressions for matrix elements of the Green function
and for the
wavefunctions. In Sec. \ref{sec:exchanges} we will derive analytical expressions
for magnetic interactions in the DE limit and address the problem
of local and global stability of the zigzag AFM ordering.
These two sections will also clarify the main
results published in Ref. \onlinecite{Solovyev1}. In Sec. \ref{sec:optics}
we will turn to the optical properties of the zigzag
chain and discuss possible implications for the resonant x-ray
scattering near Mn $K$-absorption edge.\cite{Murakami1}
In Sec. \ref{sec:LSDA} we will discuss validity of the TB picture
by comparing it with results of first-principles band structure calculations in
the local-spin-density approximation (LSDA). A brief summary will be
given in Sec. \ref{sec:summary}.

  We would like to emphasize from the very beginning that we treat the minimal
model which can be proposed for perovskite manganites. The model is based on the DE
physics and takes into account the two-fold degeneracy of the $e_g$ levels.
The combination of these two ingredients represents a necessary basis for the
analysis of manganite compounds. The main purspose of this paper is to argue that
such a simple degenerate DE model provides a very consistent description for
the low-temperature properties of the CE state in manganites.
At present it is not clear whether the same DE model will be sufficient
in the high-temperature regime.
In Sec. \ref{sec:summary}, we will collect some remarks on this matter.

  We are planning this paper as the first part in the series of two publications
dealing with the electronic structure and properties of the charge ordered
manganites. The present work contains mainly the model analysis.
Results
of the first-principles band structure calculations and the structure
optimization will be presented in Ref. \onlinecite{forthcoming}.
The first reports about this activity can be
found in Refs. \onlinecite{Lee,Terakura,Jung}. In fact, the basic idea of the
DE picture discussed in Ref. \onlinecite{Solovyev1}
was inspired by LSDA calculations of the zigzag AFM phase
in layered manganites (Ref. \onlinecite{Lee}).

\section{Electronic structure of the single zigzag chain}
\label{sec:tbmodel}

\subsection{Tight-binding Hamiltonian}
\label{subsec:hamiltonian}

  The basic translation properties of the single zigzag chain are given by
the symmetry operation $\widehat{S}_1$$=$$\{\widehat{C}(\pi)|{\bf R}_0\}$,
which combines the 180$^\circ$ rotation around the axis $[\overline{1},1,0]$,
$\widehat{C}(\pi)$, with the translation ${\bf R}_0$$=$$[\overline{a_0},a_0,0]$
connecting the sites $A$ and $A'$.
Therefore, it is convenient to choose the
atomic bases at different sites in such a form that they
can be transformed to each other by the operation
$\widehat{S}_n$$=$$\{\widehat{C}(n \pi)|n {\bf R}_0\}$, where $n$ is the integer
number of translations separating two sites.
For the $e_g$ states
this can be achieved by adopting the
following basis orbitals:
\begin{equation}
\begin{array}{l}
|1\rangle_A = -\frac{1}{2} |3z^2-r^2\rangle_A + \frac{\sqrt{3}}{2} |x^2-y^2\rangle_A \equiv |3x^2-r^2\rangle_A \\
|2\rangle_A = -\frac{\sqrt{3}}{2} |3z^2-r^2\rangle_A - \frac{1}{2} |x^2-y^2\rangle_A \equiv |y^2-z^2\rangle_A \\
\end{array}
\label{eqn:basisA}
\end{equation}
at the site $A$,
\begin{equation}
\begin{array}{l}
|1\rangle_B = |3z^2-r^2\rangle_B  \\
|2\rangle_B = |x^2-y^2\rangle_B   \\
\end{array}
\label{eqn:basisB}
\end{equation}
at the site $B$,
\begin{equation}
\begin{array}{l}
|1\rangle_{A'}=-\frac{1}{2} |3z^2-r^2\rangle_{A'}-\frac{\sqrt{3}}{2} |x^2-y^2\rangle_{A'} \equiv |3y^2-r^2\rangle_{A'} \\
|2\rangle_{A'}=-\frac{\sqrt{3}}{2} |3z^2-r^2\rangle_{A'}+\frac{1}{2} |x^2-y^2\rangle_{A'} \equiv |x^2-z^2\rangle_{A'} \\
\end{array}
\label{eqn:basisAA}
\end{equation}
at the site $A'$, and
\begin{equation}
\begin{array}{l}
|1\rangle_{B'} = |3z^2-r^2\rangle_{B'}  \\
|2\rangle_{B'} =-|x^2-y^2\rangle_{B'}   \\
\end{array}
\label{eqn:basisBB}
\end{equation}
at the site $B'$, which are periodically repeated to the whole chain.

  We assume that the kinetic hoppings between nearest-neighbor $e_g$ orbitals are
of the $dd\sigma$ type and can be parameterized in accordance with the Slater-Koster
rules.\cite{SlaterKoster} In the above basis, this yields the following nonvanishing
matrix elements between sites $A$, $B$ and $A'$ of the chain 1 in Fig. \ref{fig.chains}:
\begin{equation}
\begin{array}{l}
t^{11}_{BA}=t^{11}_{BA'}=-\frac{1}{2} \\
t^{21}_{BA}=-t^{21}_{BA'}=\frac{\sqrt{3}}{2} \\
\end{array}
\label{eqn:hoppings}
\end{equation}
In the notation $t^{LL'}_{nn'}$, $n$ and $n'$ stand for the site indices,
$L$ and $L'$ - for the orbital (basis) indices.
The absolute value of the $dd\sigma$ two-center integral is used as the energy unit.
The atomic orbitals $|2\rangle$ of the sites $A$ and $A'$ are inert and do not
participate in the hoppings. In the local representation given by
Eqs. (\ref{eqn:basisA})-(\ref{eqn:basisBB}), any translation $n{\bf R}_0$ in the
real space will transform the matrix of kinetic hoppings to itself. Therefore, we
can apply the generalized Bloch transformation $\sum_n \widehat{C}(n\pi) e^{ink}$,
$-$$\pi$$\leq$$k$$\leq$$\pi$, to the DE Hamiltonian in the real
space $\widehat{H}$$=$$\|$$-$$t^{LL'}_{nn'}\|$. This leads to the following $3$$\times$$3$
Hamiltonian in the reciprocal space, in the basis of generalized Bloch orbitals
of $|1\rangle_{A(A')}$, $|1\rangle_{B(B')}$ and $|2\rangle_{B(B')}$ type:
\begin{equation}
\widehat{H}(k)=
\left(
\begin{array}{ccc}
0 & \frac{1}{2}(1+e^{-ik}) & -\frac{\sqrt{3}}{2}(1-e^{-ik}) \\
\frac{1}{2}(1+e^{ik}) & \Delta_{\rm C} & 0 \\
-\frac{\sqrt{3}}{2}(1-e^{ik}) &  0 & \Delta_{\rm C} \\
\end{array}
\right).
\label{eqn:hamiltonian}
\end{equation}
The parameter $\Delta_{\rm C}$$>$$0$ was added in order to simulate the charge
disproportionation between the sites $A$($A'$) and $B$($B'$). The Hamiltonian has the
following eigenvalues: \\
$\varepsilon_k^\pm$$=$$\frac{1}{2}(\Delta_{\rm C}$$\pm$$
\sqrt{\Delta_{\rm C}^2+8-4\cos k})$, the bonding ($-$) and antibonding ($+$) bands,
and \\
$\varepsilon^0$$=$$\Delta_{\rm C}$, a nonbonding band, which has nonvanishing weight
only at the sites $B$ and $B'$.

  All three bands are separated by energy gaps.
Thus, if there is exactly one electron per two
sites $A$ and $B$, the low-lying bonding band is fully occupied and
the system behaves as an
one-dimensional band insulator.
Therefore, the insulating behavior of the so-called ''charge ordered'' manganites
may not be related with the charge ordering. Rather, it may be a consequence of the
peculiar zigzag AFM ordering, combined with the DE physics. Corresponding
densities of states are shown in Fig. \ref{fig.tbdos}. (Details of calculations
are given in Appendix \ref{subsec:dos}.) As expected for the one-dimensional system,
the densities of states diverge at the band edges.
Since only $3x^2$$-$$r^2$ ($3y^2$$-$$r^2$) state are occupied
at the sites $A$($A'$), the zigzag AFM ordering in the degenerate DE model
automatically leads to the orbital ordering shown in Fig. \ref{fig.chains}.

\subsection{Elements of the Green function}
\label{subsec:green}

  In the reciprocal space, the Green function for the Hamiltonian
(\ref{eqn:hamiltonian}) is given by
$\widehat{G}(k,\varepsilon)$$=$$[\varepsilon$$-$$\widehat{H}(k)$$+$$i\eta]^{-1}$,
where $\eta$ is a positive infinitesimal. It has the following matrix elements:
\begin{equation}
G^{11}_{AA}(k,\varepsilon)=\frac{1}{\varepsilon_k^+ - \varepsilon_k^-} \left\{
\frac{\varepsilon_k^+}{\varepsilon - \varepsilon_k^- +i\eta} -
\frac{\varepsilon_k^-}{\varepsilon - \varepsilon_k^+ +i\eta} \right\},
\label{eqn:g11AA}
\end{equation}
\begin{equation}
G^{11}_{BB}(k,\varepsilon)=\frac{1+\cos k}{2(\varepsilon_k^+ - \varepsilon_k^-)} \left\{
\frac{1}{\varepsilon_k^+(\varepsilon - \varepsilon_k^- +i\eta)} -
\frac{1}{\varepsilon_k^-(\varepsilon - \varepsilon_k^+ +i\eta)} \right\} +
\frac{3(1-\cos k)}{2(2-\cos k)} \frac{1}{\varepsilon - \Delta_{\rm C} + i\eta},
\label{eqn:g11BB}
\end{equation}
\begin{equation}
G^{22}_{BB}(k,\varepsilon)=\frac{3(1-\cos k)}{2(\varepsilon_k^+ - \varepsilon_k^-)} \left\{
\frac{1}{\varepsilon_k^+(\varepsilon - \varepsilon_k^- +i\eta)} -
\frac{1}{\varepsilon_k^-(\varepsilon - \varepsilon_k^+ +i\eta)} \right\} +
\frac{1+\cos k}{2(2-\cos k)} \frac{1}{\varepsilon - \Delta_{\rm C} + i\eta},
\label{eqn:g22BB}
\end{equation}
\begin{equation}
G^{11}_{AB}(k,\varepsilon)= -\frac{1+e^{-ik}}{2(\varepsilon_k^+ - \varepsilon_k^-)} \left\{
\frac{1}{\varepsilon - \varepsilon_k^- +i\eta} -
\frac{1}{\varepsilon - \varepsilon_k^+ +i\eta} \right\},
\label{eqn:g11AB}
\end{equation}
\begin{equation}
G^{12}_{AB}(k,\varepsilon)= \frac{\sqrt{3}(1-e^{-ik})}{2(\varepsilon_k^+ - \varepsilon_k^-)} \left\{
\frac{1}{\varepsilon - \varepsilon_k^- +i\eta} -
\frac{1}{\varepsilon - \varepsilon_k^+ +i\eta} \right\},
\label{eqn:g12AB}
\end{equation}
\begin{equation}
G^{12}_{BB}(k,\varepsilon)= -\frac{\sqrt{3}i \sin k}{2(\varepsilon_k^+ - \varepsilon_k^-)} \left\{
\frac{1}{\varepsilon_k^+(\varepsilon - \varepsilon_k^- +i\eta)} -
\frac{1}{\varepsilon_k^-(\varepsilon - \varepsilon_k^+ +i\eta)} +
\left( \frac{1}{\varepsilon_k^-} - \frac{1}{\varepsilon_k^+} \right)
\frac{1}{\varepsilon - \Delta_{\rm C} + i\eta} \right\}.
\label{eqn:g12BB}
\end{equation}

  The Green function elements in the real space are obtained by the Fourier transformation:
$$
G^{LL'}_{\tau \tau' + n}(\varepsilon) = \frac{1}{2\pi} \int_{-\pi}^\pi dk
e^{ink} G^{LL'}_{\tau \tau'}(k,\varepsilon),
$$
where $\tau$ and $\tau'$  correspond to either $A$ or $B$ in
Eqs. (\ref{eqn:g11AA})-(\ref{eqn:g12BB}).
This is the local representation for the Green function, meaning that the form of the
basis orbitals $L$ and $L'$ varies from site to site as prescribed by
Eqs. (\ref{eqn:basisA})-(\ref{eqn:basisBB}).

\subsection{Wave-functions}
\label{subsec:wavefunctions}

  We search the wave-functions of the Hamiltonian (\ref{eqn:hamiltonian}) in the
form:
\begin{equation}
| \Psi \rangle =
\left(
\begin{array}{c}
\cos \theta \\
\frac{1+e^{ik}}{\sqrt{2(1+\cos k})} \sin \theta \sin \phi \\
\frac{1-e^{ik}}{\sqrt{2(1-\cos k})} \sin \theta \cos \phi \\
\end{array}
\right).
\label{eqn:wf}
\end{equation}
By substituting this expression into the secular equation
$(\widehat{H}-\varepsilon)$$| \Psi \rangle$$=$$0$,
the wavefunctions can be found as follows.

  For the bonding and antibonding states $\varepsilon_k^\pm$ we obtain:
$\cot \phi_\pm$$=$$-$$\sqrt{3}\tan \frac {k}{2}$
and
$\tan \theta_\pm$$=$$\varepsilon_k^\pm/\sqrt{2-\cos k}$.
Corresponding wave-functions are given by
\begin{equation}
| \Psi_\pm \rangle_A = \cos \theta_\pm |3x^2-r^2 \rangle_A
\label{eqn:wfpmA}
\end{equation}
at the site $A$, and
\begin{equation}
| \Psi_\pm \rangle_B = \frac{\sin \theta_\pm}{\sqrt{2-\cos k}} \left\{
\frac{1}{2}(1+e^{ik})|3z^2-r^2 \rangle_B - \frac{\sqrt{3}}{2}(1-e^{ik})
|x^2-y^2 \rangle_B \right\}
\label{eqn:wfpmB}
\end{equation}
at the site $B$.

  For the nonbonding states $\varepsilon^0$$=$$\Delta_{\rm C}$ we have:
$\tan \phi_0$$=$$\sqrt{3} \tan \frac{k}{2}$
and
$\theta_0$$=$$\pi/2$.
Then, the wave-functions will have nonvanishing elements only at the corner sites.
They are given by
\begin{equation}
| \Psi_0 \rangle_B = \frac{1}{\sqrt{2-\cos k}} \left\{
\frac{\sqrt{3}}{2} (1+e^{ik}) \tan \frac{k}{2} |3z^2-r^2 \rangle_B + \frac{1}{2}
(1-e^{ik}) \tan^{-1} \frac{k}{2} |x^2-y^2 \rangle_B \right\}.
\label{eqn:wf0B}
\end{equation}

  The wave-functions at the sites $A'$($B'$) are related with the ones at the site
$A$($B$) by the transformation:
\begin{equation}
| \Psi\rangle_{A'(B')} = e^{ik} \widehat{C}(\pi) | \Psi\rangle_{A(B)}.
\label{eqn:wfAABB}
\end{equation}

\section{Magnetic interactions and stability of the CE state}
\label{sec:exchanges}

  We consider small nonuniform rotations of the spin magnetic moments on a
discrete lattice, characterized by the angles $\{ \mbox{\boldmath$\delta\varphi$}_{\it i} \}$.
If ${\bf e}^0_{\it i}$ is the direction of the spin magnetic moment at the site ${\it i}$
corresponding to an equilibrium, the new direction is given by
${\bf e}_{\it i}$$=$${\bf e}^0_{\it i}$$+$$\mbox{\boldmath$\delta$}{\bf e}_{\it i}$,
where
\begin{equation}
\mbox{\boldmath$\delta$}{\bf e}_{\it i} \equiv
\mbox{\boldmath$\delta$}^{(1)}{\bf e}_{\it i}
+ \mbox{\boldmath$\delta$}^{(2)}{\bf e}_{\it i} =
[\mbox{\boldmath$\delta\varphi$}_{\it i} \times {\bf e}^0_{\it i}] - \frac{1}{2}
(\mbox{\boldmath$\delta\varphi$}_{\it i})^2 {\bf e}^0_{\it i}.
\label{eqn:rotations}
\end{equation}
In the second order of $\{ \mbox{\boldmath$\delta\varphi$}_{\it i} \}$, the total
energy change
takes isotropic Heisenberg form
(Ref. \onlinecite{Liechtenstein}, see also Appendix \ref{apndx}):
\begin{equation}
\Delta E_t = - \frac{1}{2} \sum_{\it ij} J_{\it ij} \left\{
(\mbox{\boldmath$\delta$}^{(2)}{\bf e}_{\it i} \cdot {\bf e}^0_{\it j} ) +
(\mbox{\boldmath$\delta$}^{(1)}{\bf e}_{\it i} \cdot
\mbox{\boldmath$\delta$}^{(1)}{\bf e}_{\it j} ) +
({\bf e}^0_{\it i} \cdot \mbox{\boldmath$\delta$}^{(2)}{\bf e}_{\it j} ) \right\},
\label{eqn:heisenberg}
\end{equation}
and the parameters
$\{ J_{\it ij} \}$ can be expressed through the second derivatives of the total
energy with respect to the angles
$\{ \mbox{\boldmath$\delta\varphi$}_{\it i} \}$.\cite{Liechtenstein}
In the AFM state we have ${\bf e}^0_{\it i}$$=$${\bf e}^0_{\it j}$
(${\bf e}^0_{\it i}$$=$$-$${\bf e}^0_{\it j}$)
if the sites ${\it i}$ and ${\it j}$ belong the same (different) sublattices.

  Following the work by de Gennes (Ref. \onlinecite{deGennes})
we assume that the total energy
consists of two contributions:
$E_t[\{ {\bf e}_{\it i} \}]$$=$$E_D[\{ {\bf e}_{\it i} \}]$$+$$E_S[\{ {\bf e}_{\it i} \}]$,
where $E_D[\{ {\bf e}_{\it i} \}]$ is the kinetic (double-exchange)
energy of the $e_g$ electrons,
and
$E_S[\{ {\bf e}_{\it i} \}]$$=$$-$$\frac{1}{2} J^S \sum_{\langle {\it ij} \rangle}
{\bf e}_{\it i} \cdot {\bf e}_{\it j}$
is the energy of AFM superexchange interactions between
nearest neighbors $\langle {\it ij} \rangle$.
In practice, the coupling
$J^S$$<0$ is associated with the localized
$t_{2g}$ spins, and may also account for some interactions amongst $e_g$ electrons
beyond the DE limit.\cite{Solovyev2}
In this model, the nearest-neighbor magnetic interaction is the sum
of generally anisotropic DE coupling $J^D_{\it ij}$ (the second derivative of the
kinetic energy of the $e_g$ electrons) and the isotropic superexchange coupling:
$J_{\it ij}$$=$$J^D_{\it ij}$$+$$J^S$.

  Parameters $\{ J^D_{\it ij} \}$ depend on the magnetic state in which they are
calculated and define conditions of the local stability of this magnetic state.
The magnetic ordering is stabilized globally if it has the lowest energy amongst
all possible
locally stable magnetic states. In other words, this is the magnetic ground state
of the system.
Although the
{\it total~energy~change} near $\{ {\bf e}^0_{\it i} \}$ has
the form of the Heisenberg model -- Eq. (\ref{eqn:heisenberg}), it does not
necessary mean that the same expression is applicable for
the {\it absolute~value~of~the~total~energy} in the point $\{ {\bf e}^0_{\it i} \}$, i.e.
generally it does not hold $E_t[\{ {\bf e}^0_{\it i} \}]$$=$$
-$$\frac{1}{2}\sum_{\it j}J_{0j} ({\bf e}^0_0 \cdot {\bf e}^0_{\it j} )$.
One example illustrating such a violation is the DE model, where only
interatomic DE interactions in the FM bonds contribute to the kinetic
energy.\cite{Solovyev3}
The total energy in the DE model can still be expressed in terms of
$\{ J_{\it ij} \}$, but the form of this expression will be
different from the Heisenberg model.

\subsection{Intra-chain magnetic interactions}
\label{subsec:de_intrachain}

  In the DE limit,
magnetic interactions between nearest neighbors located in the same chain,
$J^D_{A_1B_1}$ (see Fig. \ref{fig.chains} for
notations), can be calculated as:\cite{Solovyev2}
\begin{equation}
J^D_{A_1B_1} = -\frac{1}{2\pi} {\rm Im} \int_{- \infty}^{\varepsilon_F}
d \varepsilon \left\{ G^{11}_{A_1B_1}(\varepsilon)t^{11}_{B_1A_1} +
G^{12}_{A_1B_1}(\varepsilon)t^{21}_{B_1A_1} \right\}.
\label{eqn:JA1B1.1}
\end{equation}
Using matrix elements of the Green function,
Eqs. (\ref{eqn:g11AB}) and (\ref{eqn:g12AB}), and the kinetic hoppings,
Eq. (\ref{eqn:hoppings}), we obtain:
\begin{equation}
J^D_{A_1B_1} = \frac{1}{4\pi} \int_0^\pi dk \frac {2 - \cos k}
{\sqrt{\Delta_{\rm C}^2+8 - 4 \cos k}}.
\label{eqn:JA1B1.2}
\end{equation}
The interaction $J^D_{A_1B_1}$ is always positive and responsible for
the FM coupling within the chain.
The numerical value for $\Delta_{\rm C}$$=$$0$
is
$J^D_{A_1B_1}(0)$$\simeq$$0.174$.

\begin{figure}
\centering \noindent
\resizebox{12cm}{!}{\includegraphics{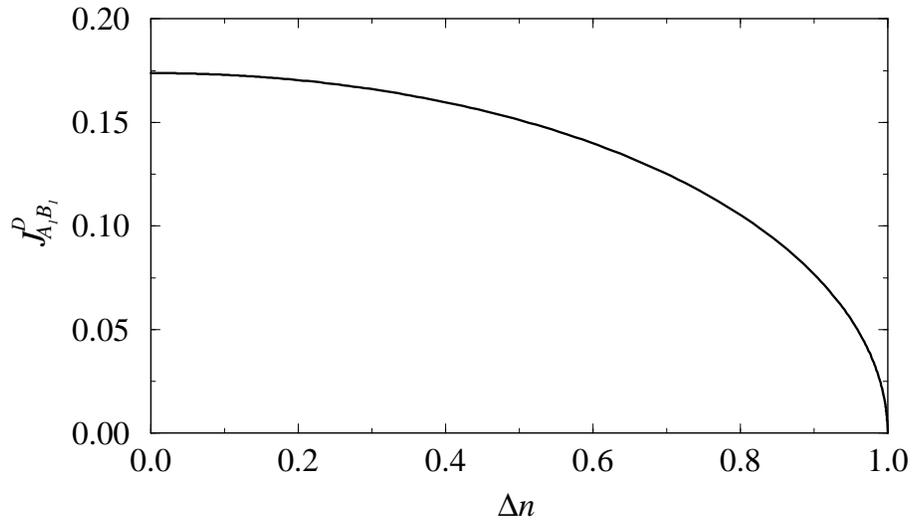}}
\caption{Effect of the charge disproportionation on the ferromagnetic
         double-exchange coupling within the chain.}
\label{fig.JA1B1}
\end{figure}

  The charge disproportionation will reduce $J^D_{A_1B_1}$.
Indeed, the former is defined as the
difference of atomic populations at the sites $A$ and $B$,
$\Delta n$$=$$n_A$$-$$n_B$, which yields:\cite{comment.7}
$$
\Delta n = \frac{1}{\pi} \int_0^{\pi}
dk \frac{\Delta_{\rm C}}{\sqrt{\Delta_{\rm C}^2+8-4\cos k}}.
$$
The dependence $J^D_{A_1B_1}$ versus $\Delta n$ is shown in Fig. \ref{fig.JA1B1}.
As $\Delta n$ increases, $J^D_{A_1B_1}$ decreases. Therefore, the charge ordering
suppresses the FM DE coupling within the chain and plays a destructive role in the
magnetic stability of the CE-type AFM ordering.

\subsection{Inter-chain magnetic interactions in the $x$-$y$ plane}
\label{subsec:de_interchain_xy}

  In the DE limit,
magnetic interactions between antiferromagnetically coupled
atoms are given by the following expression:\cite{Solovyev1,Solovyev3}
\begin{equation}
J^D_{\it ij} = \frac{1}{2\pi} {\rm Im} \int_{-\infty}^{\varepsilon_F}
d\varepsilon \sum_{\it kl} {\rm Tr}_L \left\{
\widehat{G}_{\it ik} (\varepsilon)
\widehat{t}_{\it kj} \widehat{G}_{\it jm}(\varepsilon)
\widehat{t}_{\it mi} \right\},
\label{eqn:de_AFM}
\end{equation}
where ${\rm Tr}_L$ is the trace over the orbital indices. In Eq. (\ref{eqn:de_AFM}),
the sites ${\it i}$ and ${\it k}$ belong to one zigzag chain, and the sites
${\it j}$ and ${\it m}$ belong to another zigzag chain, different from the
first one.

  If ${\it i}$$=$$A_1$ and ${\it j}$$=$$B_2$ for the geometry shown in
Fig. \ref{fig.chains}, we will have the following six nonvanishing
contributions to the exchange coupling (\ref{eqn:de_AFM}):
$$
G^{11}_{A_1A_1}(\varepsilon) t^{11}_{A_1B_2}
G^{11}_{B_2B_2}(\varepsilon) t^{11}_{B_2A_1},
$$
$$
G^{11}_{A_1A_1}(\varepsilon) t^{12}_{A_1B_2}
G^{22}_{B_2B_2}(\varepsilon) t^{21}_{B_2A_1},
$$
$$
G^{22}_{A_1A_1}(\varepsilon) t^{21}_{A_1B_2}
G^{11}_{B_2B_2}(\varepsilon) t^{12}_{B_2A_1},
$$
$$
G^{22}_{A_1A_1}(\varepsilon) t^{22}_{A_1B_2}
G^{22}_{B_2B_2}(\varepsilon) t^{22}_{B_2A_1},
$$
$$
G^{11}_{A_1A_1'}(\varepsilon) t^{11}_{A_1'B_2}
G^{11}_{B_2B_2}(\varepsilon) t^{11}_{B_2A_1},
$$
$$
G^{11}_{A_1A_1'}(\varepsilon) t^{12}_{A_1'B_2}
G^{22}_{B_2B_2}(\varepsilon) t^{21}_{B_2A_1}.
$$

  The nonbonding $x^2$$-$$z^2$ ($y^2$$-$$z^2$) orbitals of $A$ ($A'$) sites
are unoccupied and can be dropped in the analysis of magnetic interactions
of the single zigzag chain. However, they contribute to the inter-chain
interactions. The corresponding element of the Green function is
$G^{22}_{A_1A_1}(\varepsilon)$$=$$(\varepsilon$$-$$\Delta_{\rm O}
$$+$$i\eta)^{-1}$, where $\Delta_{\rm O}$ is the energy position of the
atomic $x^2$$-$$z^2$ ($y^2$$-$$z^2$) levels. The remaining elements of the Green
function listed above are given by Eqs. (\ref{eqn:g11AA})-(\ref{eqn:g22BB}).
Corresponding hopping matrix elements in the local coordinate frame
(\ref{eqn:basisA})-(\ref{eqn:basisBB}) are
$t^{11}_{A_1B_2}$$=$$t^{11}_{A_1'B_2}$$=$$\frac{1}{4}$,
$t^{12}_{A_1B_2}$$=$$-$$t^{12}_{A_1'B_2}$$=$$-$$t^{21}_{A_1B_2}$$=$$\frac{\sqrt{3}}{4}$,
and
$t^{22}_{A_1B_2}$$=$$-$$\frac{3}{4}$.
Then, noting that $\widehat{G}_{B_2B_2}(\varepsilon)$$=$$\widehat{G}_{B_1B_1}(\varepsilon)$,
we obtain the following expression:
\begin{equation}
J^D_{A_1B_2} = \frac{1}{32\pi} {\rm Im} \int_{-\infty}^{\varepsilon_F} d\varepsilon
\left\{ \left( G^{11}_{B_1B_1}(\varepsilon) + 3G^{22}_{B_1B_1}(\varepsilon) \right)
\left( G^{11}_{A_1A_1}(\varepsilon) + 3G^{22}_{A_1A_1}(\varepsilon) \right) +
\left( G^{11}_{B_1B_1}(\varepsilon) - 3G^{22}_{B_1B_1}(\varepsilon) \right)
G^{11}_{A_1A_1'}(\varepsilon) \right\}.
\label{eqn:JA1B2}
\end{equation}
The numerical value of this integral for $\Delta_{\rm O}$$=$$\Delta_{\rm C}$$=$$0$ is
$J^D_{A_1B_2}(0)$$\simeq$$0.106$, which
is significantly smaller than the value of the intra-chain
integral $J^D_{A_1B_1}(0)$. Such an anisotropy of magnetic
interactions in the DE limit readily explain the local stability of the CE-type AFM
ordering in the $x$-$y$ plane. Indeed, by combining $J^D_{A_1B_1}$ and $J^D_{A_1B_2}$
with the isotropic superexchange interaction $J^S$, we obtain the
following condition of the local stability:
\begin{equation}
J^D_{A_1B_2} < |J^S| < J^D_{A_1B_1},
\label{eqn:ls.2D}
\end{equation}
i.e., we require the total coupling $J^D_{\it ij}$$+$$J^S$ to be FM within the chain
and AFM between the chains. Since $J^D_{A_1B_2}(0)$$<$$J^D_{A_1B_1}(0)$,
this condition can be easily satisfied.

  In the two-dimensional case, this is the only
condition of the local stability of the zigzag AFM ordering.
In the three-dimensional case, the inequality (\ref{eqn:ls.2D})
shall be
combined with remaining two conditions of the local stability in the
bonds $A_1$$-$$A_3$ and $B_1$$-$$B_3$ connecting
neighboring $x$-$y$ planes -- see Fig. \ref{fig.chains}.

\subsection{Inter-chain magnetic interactions between neighboring
            $x$-$y$ planes}
\label{subsec:de_interchain_z}

  In the local basis of atomic orbitals (\ref{eqn:basisA})-(\ref{eqn:basisBB}),
the kinetic hoppings between sites $A_1$ and $A_3$, along the $z$ direction,
are
$t^{11}_{A_1A_3}$$=$$\frac{1}{4}$,
$t^{12}_{A_1A_3}$$=$$t^{21}_{A_1A_3}$$=$$\frac{\sqrt{3}}{4}$,
and
$t^{22}_{A_1A_3}$$=$$\frac{3}{4}$.
Then, applying Eq. (\ref{eqn:de_AFM}) for ${\it i}$$=$$A_1$ and ${\it j}$$=$$A_3$,
and noting that $\widehat{G}_{A_1A_1}(\varepsilon)$$=$$\widehat{G}_{A_3A_3}(\varepsilon)$,
we obtain:\cite{comment.8}
\begin{equation}
J^D_{A_1A_3} = \frac{1}{32\pi} {\rm Im} \int_{-\infty}^{\varepsilon_F} d\varepsilon
\left( G^{11}_{A_1A_1}(\varepsilon) + 6 G^{22}_{A_1A_1}(\varepsilon) \right)
G^{11}_{A_1A_1}(\varepsilon).
\label{eqn:JA1A3}
\end{equation}
The numerical value of this integral for $\Delta_{\rm O}$$=$$\Delta_{\rm C}$$=$$0$
is
$J^D_{A_1A_3}(0)$$\simeq$$0.076$.

  For the $B$-sites, the only nonvanishing hopping along the $z$-direction is
$t^{11}_{B_1B_3}$$=$$1$.
Then, applying Eq. (\ref{eqn:de_AFM}) for ${\it i}$$=$$B_1$ and ${\it j}$$=$$B_3$,
and noting that $\widehat{G}_{B_1B_1}(\varepsilon)$$=$$\widehat{G}_{B_3B_3}(\varepsilon)$,
we obtain:
\begin{equation}
J^D_{B_1B_3} = \frac{1}{2\pi} {\rm Im} \int_{-\infty}^{\varepsilon_F} d\varepsilon
\left( G^{11}_{B_1B_1}(\varepsilon) \right)^2.
\label{eqn:JB1B3}
\end{equation}
For $\Delta_{\rm O}$$=$$\Delta_{\rm C}$$=$$0$, $J^D_{B_1B_3}$ can be estimated as
$J^D_{B_1B_3}(0)$$\simeq$$0.117$, which
is different from the value reported in
Ref. \onlinecite{Solovyev1}. The reason is the following.
In Ref. \onlinecite{Solovyev1}, we employed a different basis
of atomic orbitals
at the sites $B$ and $B'$. In that basis, the Green function
$\widehat{G}_{B_1B_1}(\varepsilon)$ has off-diagonal elements with respect to the
orbital indices, which have been neglected in the analysis of magnetic interactions
in Ref. \onlinecite{Solovyev1}. The new values of the exchange integrals take
into account these terms. The correction appears to be small for the
in-plane exchange interaction $J^D_{A_1B_2}(0)$ and modifies it only in the fourth
digit after the comma.
However, the change of the inter-plane exchange interaction
$J^D_{B_1B_3}(0)$ appears to be more significant. With this new value of
$J^D_{B_1B_3}(0)$, the main statement of
Ref. \onlinecite{Solovyev1} appears to be even stronger. If $J^S$ satisfies the
condition
\begin{equation}
\max \left\{ J^D_{A_1B_2}, J^D_{A_1A_3}, J^D_{B_1B_3} \right\}  < |J^S| < J^D_{A_1B_1},
\label{eqn:ls.3D.hard}
\end{equation}
the FM coupling is stabilized in the bond $A_1$-$B_1$, and the AFM
coupling is stabilized in the bonds $A_1$-$B_2$, $A_1$-$A_3$ and $B_1$-$B_3$.
Since $J^D_{B_1B_3}(0)$$\equiv$$
\max \left\{ J^D_{A_1B_2}(0), J^D_{A_1A_3}(0), J^D_{B_1B_3}(0) \right\}
$$<$$J^D_{A_1B_1}(0)$,
appearance of the three-dimensional
CE-type AFM ordering can be fully explained by
the anisotropy of magnetic interactions in the DE limit.
We will call Eq. (\ref{eqn:ls.3D.hard}) the ''hard'' condition of the local stability.
Similar to Ref. \onlinecite{Solovyev1}, we can also consider the CE-type AFM
ordering in the regime:
\begin{equation}
J^D_{A_1B_2} < |J^S| < J^D_{B_1B_3},
\label{eqn:ls.3D.soft}
\end{equation}
when the total coupling in the bond $B_1$-$B_3$ is ferromagnetic. Even in this case,
the CE-type AFM ordering may remain stable due to the joint effect of
magnetic interactions in the bonds $A_1$-$A_3$, $A_1$-$B_1$ and $A_1$-$B_2$.
We will call Eq. (\ref{eqn:ls.3D.soft}) the ''soft'' condition of the local stability.

\subsection{Local and global stability of the zigzag antiferromagnetic
            ordering}
\label{subsec:local-global}

  In Ref. \onlinecite{Solovyev3} we argued that in the DE limit
the kinetic energy for a collinear magnetic state M
is related with
the interatomic DE coupling parameter in the FM bonds, $J_{\uparrow \uparrow}^D({\rm M})$, as
$E_D({\rm M})$$=$$-$$2z_{\uparrow \uparrow} J_{\uparrow \uparrow}^D({\rm M})$,
where $z_{\uparrow \uparrow}$ is the number of FM bonds.
The superexchange energy is given by
$E_S({\rm M})$$=$$-$$\frac{1}{2} J^S (z_{\uparrow \uparrow}$$-$$z_{\uparrow \downarrow})$,
where $z_{\uparrow \downarrow}$ is the number of AFM bonds.
This represents the connection between
the parameters $\{ J_{\it ij}^D,J^S \}$ and the total energy of the DE model.

  Let us consider the three-dimensional case. For the hole concentration
$x$$=$$0.5$ we can expect three possibilities: the FM ordering
(${\rm M}$$=$${\rm F}$, $z_{\uparrow \uparrow}$$=$$6$, and
$z_{\uparrow \downarrow}$$=$$0$),
the A-type AFM ordering (${\rm M}$$=$${\rm A}$, $z_{\uparrow \uparrow}$$=$$4$,
and $z_{\uparrow \downarrow}$$=$$2$),
and the CE-type AFM ordering (${\rm M}$$=$${\rm CE}$,$z_{\uparrow \uparrow}$$=$$2$, and
$z_{\uparrow \downarrow}$$=$$4$).
The chain-like C-type AFM ordering is unstable for $x$$=$$0.5$,\cite{Solovyev3}
and can be excluded from the analysis. Corresponding total energies are given by:
$E_t({\rm F})$$=$$-$$12 J^D_{\uparrow \uparrow}({\rm F})$$-$$3J^S$,
$E_t({\rm A})$$=$$-8$$J^D_{\uparrow \uparrow}({\rm A})$$-$$J^S$,
and
$E_t({\rm CE})$$=$$-$$4 J^D_{A_1B_1}$$+$$J^S$.

  Conditions of the local stability of the FM and A-type AFM states are
$|J^S|$$<$$J^D_{\uparrow \uparrow}({\rm F})$
and
$J^D_{\uparrow \downarrow}({\rm A})$$<$$|J^S|$$<J^D_{\uparrow \uparrow}({\rm A})$,
respectively,
where $J^D_{\uparrow \downarrow}({\rm A})$
is the nearest-neighbor
exchange interaction between antiferromagnetically coupled atoms
in the DE limit.
''Hard'' and ''soft'' conditions of the local stability of the CE-type AFM
ordering
are given by Eqs. (\ref{eqn:ls.3D.hard}) and (\ref{eqn:ls.3D.soft}),
respectively.

\begin{figure}
\centering \noindent
\resizebox{12cm}{!}{\includegraphics{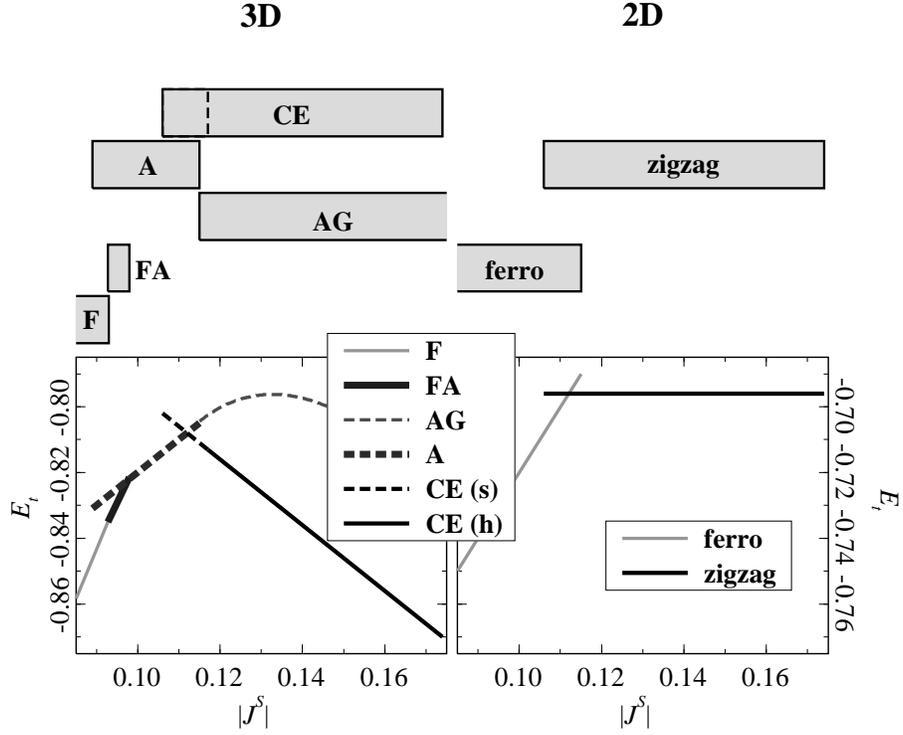}}
\caption{Total energies (bottom) and the regions of local stability (top)
         for various magnetic states in the three-dimensional (3D) and
         two-dimensional (2D) perovskite lattices for $x$$=$$0.5$.
         The data for the CE-type antiferromagnetic ordering are divided
         in two parts corresponding to the hard (h) and soft (s) regimes
         of the local stability.}
\label{fig.local-global}
\end{figure}

  Numerical values of the exchange integrals for $x$$=$$0.5$ and
$\Delta_{\rm C}$$=$$\Delta_{\rm O}$$=$$0$ are
$J^D_{\uparrow \uparrow}({\rm F})$$\simeq$$0.093$, $J^D_{\uparrow \uparrow}({\rm A})$$
\simeq$$0.115$ and $J^D_{\uparrow \downarrow}({\rm A})$$\simeq
$$0.089$.\cite{Solovyev2,Solovyev3,comment.1}
Corresponding phase diagram is shown
in Fig. \ref{fig.local-global}, where we plot the total energies for each magnetic state
on the interval inside of which this magnetic state is locally stable.
We also added the data for two canted spin states which may exist for
$x$$=$$0.5$. These solutions of the DE model
can be regarded as intermediate states
between:\\
(1) the FM and A-type AFM ordering (the canted spin state of
the FA type);\\
(2) the A- and G-type AFM ordering (the canted spin state of
the AG type).\cite{comment.6}

  From Fig. \ref{fig.local-global} we can conclude that the CE-type
AFM ordering may coexist only with the A-type AFM
ordering
and the canted spin ordering of the AG type. The FM ordering and the
canted spin ordering of the FA type may develop only for smaller
$|J^S|$, when the CE-type AFM ordering is already unstable.
If $|J^S|$$>$$0.112$ (the crossing point of CE and A total energy curves),
the CE-type AFM state has
the lowest energy amongst the states CE, A and AG, and is the magnetic ground
state of the system.

  In the two-dimensional case, the phase diagram is very simple: the total
energy of the zigzag AFM state ($z_{\uparrow \uparrow}$$=$$2$,
$z_{\uparrow \downarrow}$$=$$2$),
$E_t({\rm zigzag})$$=$$-$$4 J^D_{A_1B_1}$,
shall be compared with the total energy of the two-dimensional FM
state ($z_{\uparrow \uparrow}$$=$$4$,
$z_{\uparrow \downarrow}$$=$$0$),
$E_t({\rm ferro})$$=$$-$$8 J^D_{\uparrow \uparrow}({\rm ferro})$$-$$2J^S$,
where $J^D_{\uparrow \uparrow}({\rm ferro})$$\simeq$$0.115$.\cite{Solovyev3}
Conditions of the local stability
are:
$|J^S|$$<$$J^D_{\uparrow \uparrow}({\rm ferro})$
for the two-dimensional FM ordering, and
Eq. (\ref{eqn:ls.2D}) for the zigzag AFM ordering. Two states
coexist in the interval
$J^D_{A_1B_2}$$<$$|J^S|$$<$$J^D_{\uparrow \uparrow}({\rm ferro})$.
If $|J^S|$$>$$0.112$, the
zigzag AFM state has lower energy than the FM state
and vice versa.

\subsection{Simple model for the Jahn-Teller distortion around the bridge
            sites}
\label{subsec:JT.exchange}

  We assume that the Jahn-Teller distortion (JTD) results in the shift of
atomic $y^2$$-$$z^2$/$3x^2$$-$$r^2$ ($x^2$$-$$z^2$/$3y^2$$-$$r^2$) levels
of the $A$ ($A'$) sites by $\pm$$\frac{1}{2}\Delta_{\rm JT}$. This is
equivalent to the following choice of the parameters of our original
Hamiltonian: $\Delta_{\rm O}$$=$$\Delta_{\rm JT}$ and $\Delta_{\rm C}$$=
$$\frac{1}{2}\Delta_{\rm JT}$, i.e. in addition to the local splitting at
the $A$ ($A'$) sites, we shift all the states upwards by $\frac{1}{2}\Delta_{\rm JT}$.
Obviously, the constant shift does not affect the magnetic
interactions.

\begin{figure}
\centering \noindent
\resizebox{11cm}{!}{\includegraphics{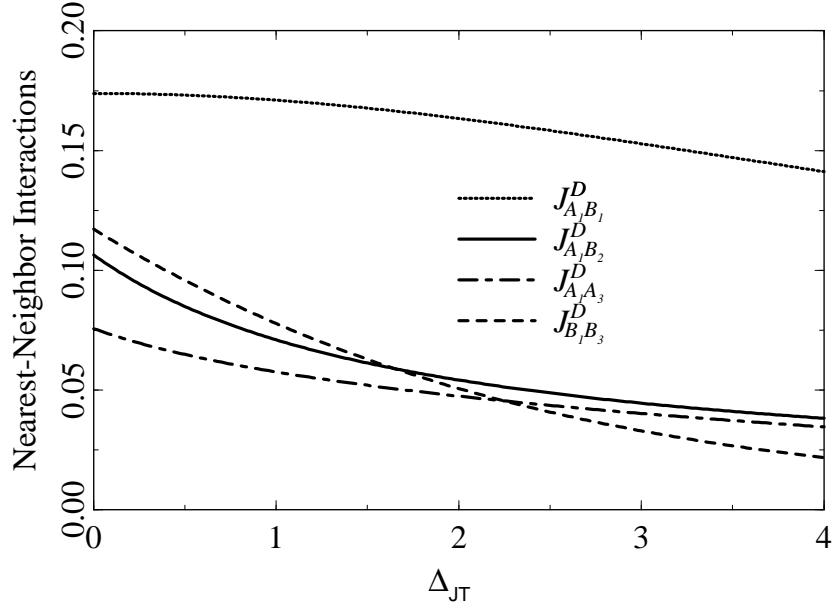}}
\caption{Nearest-neighbor interactions in the double-exchange limit as the function
         of the Jahn-Teller distortion at the bridge sites.}
\label{fig.JT.exchange}
\end{figure}

  Thus, the JTD in such a picture will be accompanied by
a charge disproportionation,
and suppress the FM DE coupling within the chain
(see Fig. \ref{fig.JA1B1}). This is, in principle, a negative consequence which
lowers the upper boundary of the local stability of the zigzag AFM
ordering.
Nevertheless, for relatively small distortions, the change of $J^D_{A_1B_1}$ is also
expected to be
small. The effect of $\Delta_{\rm JT}$ on the interatomic interactions in
the AFM bonds
$A_1$-$B_2$, $A_1$-$A_3$ and $B_1$-$B_3$ is more dramatic. The results are
summarized in Fig. \ref{fig.JT.exchange}.

  In order to evaluate the magnitude of the oxygen displacement which results
in the splitting $\Delta_{\rm JT}$, we consider the following procedure:\cite{Solovyev4}
(i) we start with a general TB Hamiltonian on an undistorted
cubic lattice, in the basis of Mn($3de_g$) and O($2p$) orbitals interacting
via nearest-neighbor $pd\sigma$ hoppings, the distance-dependence of which is
simulated by the Harrison low: $(pd\sigma)$$\propto$$d^{-7/2}$; (ii) we switch on
the JTD of the MnO$_6$ octahedra around the bridge sites,
the $Q_3$ mode,\cite{Kanamori} (it is assumed that the total volume of the MnO$_6$
octahedron remains to be constant); (iii) we project out the O($2p$) states.
This results in the on-site splitting of $e_g$ orbitals,
$\Delta_{\rm JT}$$\simeq$$28 \delta$
(in units of the effective $dd\sigma$ two-center
integral for the undistorted cubic lattice), where the oxygen displacement is
described by the parameter $\delta$$=$$(d_L$$-$$d_S)/(d_L$$+$$d_S)$,
with $d_L$ ($d_S$) being the long (short) Mn-O bond length in the
distorted MnO$_6$ octahedron.

  Then, the behavior of magnetic interactions shown in Fig. \ref{fig.JT.exchange}
suggests that even a modest distortion $\delta$$=$$0.02$
($\Delta_{\rm JT}$$\simeq$$0.56$, see also Ref. \onlinecite{comment.3})
may reduce the FM contribution to the
AFM bonds $A_1$-$B_2$ and $B_1$-$B_3$ by up to 20\%, whereas
the coupling in the FM bond $A_1$-$B_1$ remains practically
unchanged. Therefore, the JTD significantly lowers the
low boundary of the local stability for the zigzag AFM ordering.
As the result, the range of parameters $|J^S|$ which stabilize the zigzag
AFM ordering
significantly widens.

%
%

\section{Optical properties of the zigzag chain}
\label{sec:optics}

  In order to evaluate the optical conductivity tensor
$\widehat{\sigma}(\omega)$ for the single zigzag chain we take the following steps.
First, we include the
Mn($4p$) states in our model and evaluate their admixture into the eigenfunctions
of Hamiltonian (\ref{eqn:hamiltonian}) using the simplest perturbation
theory. We assume that the main perturbation is caused by the kinetic hoppings
between nearest-neighbor Mn($4p$) and Mn($3de_g$) orbitals.\cite{comment.4}
The hoppings are of the $pd\sigma$ type and can be parameterized in accordance
with Slater-Koster rules.\cite{SlaterKoster} It is further assumed that the
energy splitting between the atomic $4p$ and $3de_g$ levels, $\Delta_{4p-e_g}$,
is much larger than the corresponding bandwidths, so that the band dispersion
can be neglected in the perturbation theory expansion (this is certainly
a very crude approximation, especially for the $4p$ band, the validity of which
will be discussed in Sec. \ref{sec:LSDA}). Thus, the mixture of the
Mn($4p$) and Mn($3de_g$) states will be controlled by the parameter
$\alpha$$=$$(pd\sigma)_{4p-e_g}$$/$$\Delta_{4p-e_g}$, where
$(pd\sigma)_{4p-e_g}$ is the $4p$-$3de_g$
two-center integral.\cite{SlaterKoster}
Then, by knowing distribution of the Mn($4p$) and Mn($3de_g$) states
we can find matrix elements of the interband optical transitions
$\varepsilon^-_k$$\rightarrow$$\varepsilon^0$
and $\varepsilon^-_k$$\rightarrow$$\varepsilon^+_k$,
and express them through the intra-atomic dipole
$4p$-$3de_g$
matrix elements.
Finally, the optical conductivity can be calculated using the Kubo formula. Due to
the one-dimensional character of the problem, the solution along this line
can be carried out analytically.
For simplicity we will consider only the case of homogeneous charge distribution
($\Delta_{\rm C}$$=$$0$).

 The simple TB picture
considered in this section
is of course very oversimplified and cannot provide
a good quantitative description.
However, it appears to be very useful in the analysis of symmetry properties
and the structure of the conductivity tensor in terms of partial contributions
from different sites of the system.

\subsection{Distribution of Mn($4p$) states}
\label{subsec:Mn.4p}

  First, we consider the admixture of the $4p$-orbitals $|x\rangle$ and
$|y\rangle$ into the bonding and antibonding states of the Hamiltonian (\ref{eqn:hamiltonian}).
Taking in to account the explicit form of the wavefunctions at the sites $A$ and $A'$
given by
Eqs. (\ref{eqn:wfpmA}) and (\ref{eqn:wfAABB}), and considering the nearest-neighbor
$pd\sigma$ hoppings from the sites $A$ and $A'$ to the site $B$ of the chain 1
in Fig. \ref{fig.chains}, we obtain the following corrections to the wavefunctions
at the site $B$, in the first order of $\alpha$:
\begin{equation}
|\delta \Psi_\pm \rangle_B =
- \alpha \cos \theta_\pm \left\{ |x\rangle_B + e^{ik} |y\rangle_B \right\}.
\label{eqn:wfpmBp}
\end{equation}
Starting with the sites $B$ and $B'$ and considering the hoppings onto the
site $A$, we obtain:
\begin{equation}
|\delta \Psi_\pm \rangle_A =
\frac{i \alpha \sin \theta_\pm \sin k}{\sqrt{2 - \cos k}} |x\rangle_A.
\label{eqn:wfpmAp}
\end{equation}
Similar analysis applied to the nonbonding state $|\Psi_0\rangle$ yields:
\begin{equation}
|\delta \Psi_0 \rangle_A =
\sqrt{3} i \alpha
\frac{\cos k}{\sqrt{2 - \cos k}} |x\rangle_A.
\label{eqn:wf0Ap}
\end{equation}
Since $|\Psi_0\rangle$ has nonvanishing elements only at the sites of $B$ and $B'$
type and the kinetic $3de_g$-$4p$ hoppings are restricted by the nearest neighbors,
the correction $|\delta \Psi_0 \rangle$ may exist only at the (neighboring)
sites of $A$ type.

  Corresponding corrections at the sites $A'$ and $B'$ can be found
using Eq. (\ref{eqn:wfAABB}).

\subsection{Matrix elements of the optical transitions}
\label{subsec:me.optics}

  The nonvanishing intra-atomic dipole matrix elements obey the following rules:\cite{Slater}
$\langle 3x^2$$-$$r^2 |\widehat{x}|x \rangle$$ = $$
\langle 3y^2$$-$$r^2 |\widehat{y}|y \rangle $$\equiv$$ \beta$,
$\langle 3z^2$$-$$r^2 |\widehat{x}|x \rangle$$ = $$
\langle 3z^2$$-$$r^2 |\widehat{y}|y \rangle$$=$$-$$\frac{1}{2} \beta$,
and
$\langle x^2$$-$$y^2 |\widehat{x}|x \rangle$$ = $$
- $$\langle x^2$$-$$y^2 |\widehat{y}|y \rangle$$ = $$\frac{\sqrt{3}}{2} \beta$.
Then, using results of Secs.
\ref{subsec:wavefunctions} and \ref{subsec:Mn.4p} for $\Delta_{\rm C}$$=$$0$, we
obtain the following
matrix elements:
\begin{equation}
\langle \Psi_- | \widehat{x} | \delta \Psi_0 \rangle_A =
\langle \Psi_- | \widehat{y} | \delta \Psi_0 \rangle_{A'}^\ast =
-\sqrt{\frac{3}{2}} i \alpha \beta \frac{ \cos k}{\sqrt{2 - \cos k}}
\label{eqn:otbnA}
\end{equation}
for the bonding-nonbonding transitions at the sites $A$ and $A'$;
\begin{equation}
\langle \delta \Psi_- | \widehat{x} | \Psi_0 \rangle_B =
\langle \delta \Psi_- | \widehat{y} | \delta \Psi_0 \rangle_B^\ast =
\langle \delta \Psi_- | \widehat{x} | \Psi_0 \rangle_{B'} =
\langle \delta \Psi_- | \widehat{y} | \delta \Psi_0 \rangle_{B'}^\ast =
-\sqrt{\frac{3}{8}} i \alpha \beta \frac{1}{\sqrt{2 - \cos k}}
\label{eqn:otbnB}
\end{equation}
for the bonding-nonbonding transitions at the sites $B$ and $B'$;
\begin{equation}
\langle \delta \Psi_- | \widehat{x} | \Psi_+ \rangle_A +
\langle \Psi_- | \widehat{x} | \delta \Psi_+ \rangle_A =
 \langle \delta \Psi_- | \widehat{y} | \Psi_+ \rangle_{A'}^\ast
+ \langle \Psi_- | \widehat{y} | \delta \Psi_+ \rangle_{A'}^\ast =
- i \alpha \beta \frac{ \sin k}{\sqrt{2 - \cos k}}
\label{eqn:otbaA}
\end{equation}
for the bonding-antibonding transitions at the sites $A$ and $A'$; and
\begin{equation}
\langle \delta \Psi_- | \widehat{x} | \Psi_+ \rangle_B +
\langle \Psi_- | \widehat{x} | \delta \Psi_+ \rangle_B =
\langle \delta \Psi_- | \widehat{y} | \Psi_+ \rangle_B^\ast +
\langle \Psi_- | \widehat{y} | \delta \Psi_+ \rangle_B^\ast =
 \langle \delta \Psi_- | \widehat{x} | \Psi_+ \rangle_{B'}
+ \langle \Psi_- | \widehat{x} | \delta \Psi_+ \rangle_{B'} =
 \langle \delta \Psi_- | \widehat{y} | \Psi_+ \rangle_{B'}^\ast
+ \langle \Psi_- | \widehat{y} | \delta \Psi_+ \rangle_{B'}^\ast =
\frac{1}{2} i \alpha \beta
\frac{\sin k}{\sqrt{2 - \cos k}}
\label{eqn:otbaB.1}
\end{equation}
for the bonding-antibonding transitions at the sites $B$ and $B'$.

\subsection{Optical conductivity}
\label{subsec:cunductivity}

  The inter-band optical conductivity tensor can be obtained using the Kubo
formula:
\begin{equation}
\sigma_{\gamma \gamma'} (\omega) \propto \frac{1}{\omega}
\int_0^\pi dk \langle \Psi_{\rm i} | \widehat{\gamma} | \Psi_{\rm f} \rangle
\langle \Psi_{\rm f} | \widehat{\gamma}' | \Psi_{\rm i} \rangle
\delta(\omega - \varepsilon_k^{\rm f} + \varepsilon_k^{\rm i}),
\label{eqn:Kubo}
\end{equation}
where $\gamma(\gamma')$$=$ $x$, $y$; $\Psi_{\rm i}$ and $\varepsilon_k^{\rm i}$ stand
for the initial state ${\rm i}$$=$''$-$'';
$\Psi_{\rm f}$ and $\varepsilon_k^{\rm f}$ stand
for the final state ${\rm f}$$=$''$0$'' or ''$+$''.
We consider only site-diagonal contributions in Eq. (\ref{eqn:Kubo}).
The integral can be evaluated along the same line as for the local densities of states
in Appendix \ref{subsec:dos}.
Then, we obtain the following nonvanishing contributions to the conductivity tensor:
\begin{equation}
\sigma_{xx}^A(\omega) = \sigma_{yy}^{A'}(\omega)
\label{eqn:ocAsym}
\end{equation}
at the sites $A$ and $A'$,
where
\begin{equation}
\sigma_{xx}^A(\omega) \propto
\frac{3}{\omega^2} \frac{(2-\omega^2)^2}
{\sqrt{(3 - \omega^2)(\omega^2 - 1)}}
\label{eqn:ocAbn}
\end{equation}
for the bonding-nonbonding transitions (we drop the common multiplier
$\alpha^2 \beta^2$), and
\begin{equation}
\sigma_{xx}^A(\omega) \propto
\frac{1}{2 \omega^2}
\sqrt{(12 - \omega^2)(\omega^2 - 4)}
\label{eqn:ocAba}
\end{equation}
for the bonding-antibonding transitions;
\begin{equation}
\sigma_{xx}^B(\omega) = - \sigma_{xy}^B(\omega) =
- \sigma_{yx}^B(\omega) = \sigma_{yy}^B(\omega)
\label{eqn:ocBsym.1}
\end{equation}
and
\begin{equation}
\widehat{\sigma}^{B'}(\omega) = \widehat{\sigma}^B(\omega),
\label{eqn:ocBsym.2}
\end{equation}
at the sites $B$ and $B'$,
where
\begin{equation}
\sigma_{xx}^B(\omega) \propto
\frac{3}{4 \omega^2} \frac{1}
{\sqrt{(3 - \omega^2)(\omega^2 - 1)}}
\label{eqn:ocBbn}
\end{equation}
for the bonding-nonbonding transitions, and
\begin{equation}
\sigma_{xx}^B(\omega) \propto
\frac{1}{8 \omega^2}
\sqrt{(12 - \omega^2)(\omega^2 - 4)}
\label{eqn:ocBba}
\end{equation}
for the bonding-antibonding transitions.

  The results are shown in Fig. \ref{fig.opt.tb}.
The conductivity spectrum consists of two parts originating from the
bonding-nonbonding and bonding-antibonding transitions, and spreading
within the intervals $1$$\leq$$\omega$$\leq$$\sqrt{3}$ and
$2$$\leq$$\omega$$\leq$$2\sqrt{3}$, respectively.
Taking into account that a realistic estimate for the effective $dd\sigma$
two-center integral is about 0.7 eV (for the undistorted lattice, without
buckling of Mn-O-Mn bonds),\cite{Solovyev2}
they roughly correspond to the energy ranges
$0.7$$\leq$$\omega$$\leq$$1.2$ eV and
$1.4$$\leq$$\omega$$\leq$$2.4$ eV.
We note that
the components of the conductivity tensor
associated with
the bridge ($A$,$A'$) and the corner
($B$,$B'$) sites are characterized by different optical axes. The tensors
$\widehat{\sigma}^A(\omega)$ and $\widehat{\sigma}^{A'}(\omega)$ are
diagonal in the basis of cubic axes $x$$=$$[1,0,0]$ and $y$$=$$[0,1,0]$
shown in Fig. \ref{fig.chains}.
In this basis, each of the tensors has only one nonvanishing diagonal
element, correspondingly $\sigma_{xx}^A(\omega)$ and $\sigma_{yy}^{A'}(\omega)$,
connected by the symmetry transformation (\ref{eqn:ocAsym}).
Therefore, the total conductivity $\widehat{\sigma}^A(\omega)$$+$$
\widehat{\sigma}^{A'}(\omega)$
behaves as an isotropic object in the
plane $x$-$y$. Conversely, the tensors
$\widehat{\sigma}^B(\omega)$$=$$\widehat{\sigma}^{B'}(\omega)$
are diagonal in the basis of orthorhombic axes
$a$$=$$[1,1,0]$ and $b$$=$$[\overline{1},1,0]$,
which are rotated with respect to cubic ones by 45$^\circ$.
In the orthorhombic frame, they have only one nonvanishing diagonal element
$\sigma_{bb}^B(\omega)$$=$$\sigma_{bb}^{B'}(\omega)$$=$$2\sigma_{xx}^B(\omega)$
in the direction of propagation of the zigzag chain (the $b$-direction).
This gives rise to an anisotropy of the optical conductivity, which
has been observed experimentally in
La$_{1/2}$Sr$_{3/2}$MnO$_4$ by using a birefringence technique.\cite{Ishikawa}

  The energy range corresponding to the bonding-nonbonding transitions roughly
agrees with the position of the mid-infrared peak observed in
La$_{1/2}$Sr$_{3/2}$MnO$_4$ around 1.0-1.3 eV.\cite{Jung,Ishikawa}
High intensity of experimental peak may be partly attributed to the
(quasi-) one-dimensional character of the problem and divergence
of the optical conductivity
at the band edges.
Another possibility is
the lack of the inversion symmetry at the corner sites $B$ and $B'$. In this case,
the atomic $3de_g$ and $4p$ orbitals may belong to the same representation of the
local symmetry group. This is an additional channel of mixing of the $3de_g$ and
$4p$ states at the corner atoms, which
contributes to probabilities of the optical transitions.

  The JTD around the bridge sites can affect the conductivity
spectrum indirectly, through the induced charge disproportionation, in a way similar
to the intra-chain DE interaction considered in Sec. \ref{subsec:JT.exchange}.
Since the charge disproportionation $\Delta_{\rm C}$$>$$0$
increases the energy gap
between the bonding and nonbonding
bands and decrease the width of the former,
we expect
the upward shift of
the optical absorption center
and narrowing of the spectrum
in the region of the bonding-nonbonding transitions.

  If the charge disproportionation does not take place ($\Delta_{\rm C}$$=$$0$), the
optical conductivity is expected to be
the smooth function in the region of bonding-antibonding
transitions and reveals the same symmetry properties as
for the bonding-nonbonding ones.

\subsection{Resonant x-ray scattering near the Mn $K$-absorption edge}
\label{subsec:MnK}

  The resonant x-ray scattering near the Mn $K$-absorption edge is considered
as a powerful experimental tool, which allows to probe the charge and orbital
distribution in the perovskite manganites.\cite{Murakami1,Murakami2}
The interpretation of experimental data is, however, hampered by the fact
that the direct dipole transitions are allowed to the
$4p$ states but not directly to the $3d$ states in the
case of $K$-absorption.
Therefore, the understanding of how the
$4p$ states interact with polarized $3de_g$ orbitals represents a special
interest.\cite{Ishihara,Elfimov}
Here we would like to show how the geometry of the zigzag chain imposes some
constraints on the distribution of the Mn($4p$) states, which should be reflected
in the lowest energy part of the $K$-absorption spectrum of manganites with
the zigzag AFM ordering.

  In accordance with the TB picture (Fig. \ref{fig.tbdos}), the first
unoccupied band which is involved in the transitions is the nonbonding band composed
of the $3de_g$ orbitals of the corner atoms $B$ and $B'$.
If the hoppings between $3de_g$ and
$4p$ orbitals are restricted by nearest neighbors, the $4p$ states
will have nonvanishing weight only at the bridge sites $A$ and $A'$
in the nonbonding part of the spectrum.
This interaction, given by Eqs. (\ref{eqn:wf0Ap}) and (\ref{eqn:wfAABB}), results
in the ordering of $|x\rangle_A$ and $|y\rangle_{A'}$ orbitals.
Since the $e_g$ orbitals of the bridge atoms do not contribute to the nonbonding
band, there is nothing to re-expand over the $4p$ orbitals of the neighboring
(corner) atoms, and the corresponding weight of the $4p$ states at the
corner atoms will be zero.

  Thus, in accordance with our TB picture,
the lowest energy part of the $K$-absorption spectrum probes the
$|x\rangle_A$ and $|y\rangle_{A'}$ states of
the bridge atoms, which carry the information about the distribution of the
$e_g$ states at the corner atoms.
%

\section{Comparison with LSDA band structure calculations}
\label{sec:LSDA}

  The purpose of this section is to discuss validity of the TB
model for the zigzag AFM state
by comparing it with results of first-principles
band structure calculations in the
local-spin-density approximation (LSDA).
LSDA and its extension in the form of the generalized gradient approximation
(GGA) work reasonably well in
manganites.\cite{Solovyev2,LaMnO3.1,Solovyev7,Sawada,review.2,Fang}
There are several unresolved problems,
such as the relative stability of the
Jahn-Teller distorted insulating A-type AFM state and the metallic FM state in LaMnO$_3$:
the standard band calculations account well for various magnetic
properties of LaMnO$_3$ with the experimental crystal structure,\cite{LaMnO3.1,Solovyev7}
however the full optimization of the crystal structure substantially reduces the
JTD and makes the A-type AFM state energetically less
favorable than the FM state.\cite{Sawada}
To say the truth, similar problem was encountered also in La$_{1/2}$Sr$_{3/2}$MnO$_4$,
where
after the structure optimization
in GGA
the FM state had lower energy than the zigzag AFM state.\cite{forthcoming}
However, it is also true that at present the problem has no feasible solution.
For example, the LDA$+$$U$ (or GGA$+$$U$) approach only worsen the picture of
magnetic interactions in LaMnO$_3$.\cite{Solovyev5}
This may be related with the fact that the relative position of the oxygen
$2p$ and Mn($3d$) states (the so-called charge-transfer energy) is incorrect
in LDA$+$$U$, that
leads to a large error
in the calculated parameters of magnetic interactions.\cite{Solovyev6}
Thus, we believe that the LSDA provides the most reliable reference point
for the analysis of magnetic properties of manganites.

   The quasi-one-dimensional TB model
is a substantial simplification for realistic material,
in several respect. (i) The DE limit is not realized in
perovskite manganites.\cite{Solovyev2}
So, in the zigzag AFM state the chains are not fully isolated from each other,
and
the system is not
strictly one-dimensional.
(ii) The distribution of Mn($3de_g$) states is affected by other factors, which are not
included explicitely in the TB model, particularly by the oxygen $2p$
and Mn($3dt_{2g}$) states.\cite{Mahadevan}
(iii) The Mn($4p$) states
generally form a very broad band. In this respect, our attempt to
simulate the optical properties by starting with atomic $4p$-levels and considering
their interations with the $e_g$ bands in a perturbative manner
was a crude approximation. It allows
to catch the basic symmetry properties of the optical
conductivity tensor, but certainly
not all the details and the absolute values of the conductivity itself.

  We use the ASA-LMTO method in the nearly orthogonal
representation.\cite{Andersen,LMTO}
The Wigner-Seitz sphere radii for nonequivalent sites were chosen from the charge
neutrality condition inside the spheres. The calculations have been performed
for the phenomenological virtual-crystal alloy
Y$_{1/2}$Sr$_{3/2}$MnO$_4$
using the lattice parameters of the real two-dimensional perovskite
La$_{1/2}$Sr$_{3/2}$MnO$_4$.\cite{Sternlieb,Murakami1}
In addition, we also investigated the effect of the
JTD around the bridge sites.
The latter is characterized by the parameter $\delta$ introduced in Sec. \ref{subsec:JT.exchange}.

\begin{figure}
\centering \noindent
\resizebox{10cm}{!}{\includegraphics{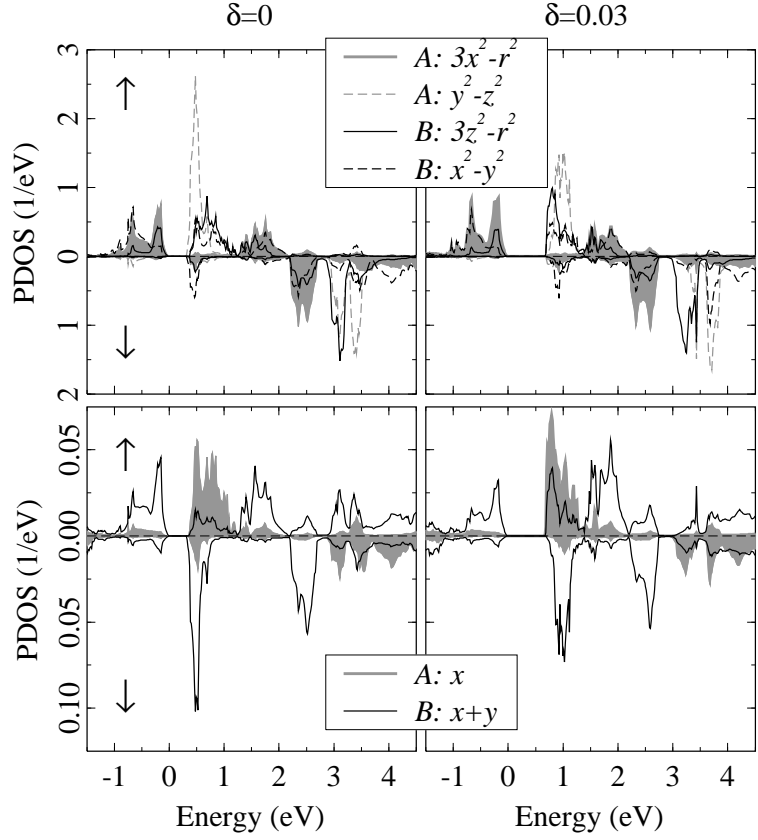}}
\caption{Partial densities of Mn($3de_g$) (top) and Mn($4p$) (bottom) states
         for Y$_{1/2}$Sr$_{3/2}$MnO$_4$ in the zigzag antiferromagnetic state.
         Two panels show results of calculations with (right) and without (left)
         the Jahn-Teller distortion.
         The Fermi level (the top of occupied band) is at zero.
         The arrows $\uparrow$ and $\downarrow$ stand for the majority- and
         minority-spin states.}
\label{fig.LSDAdos}
\end{figure}

  The partial densities of Mn($3de_g$) and Mn($4p$) states are shown in Fig. \ref{fig.LSDAdos}.
They reveal many similarities to the TB picture. There are aslo some
differences. We note the following. First, there is the clear splitting
of the majority-spin
$3de_g$ states into the bonding band located around $-$$0.5$ eV (the case of
$\delta$$=$$0$), the nonbonding bands around $0.7$ eV, and the antibonding band
around $1.7$ eV. Two nonbonding bands are composed of the $e_g$ states of
the corner ($B$ and $B'$) atoms and the $y^2$$-$$z^2$ ($x^2$$-$$z^2$) states of the
bridge $A$ ($A'$) atoms. Even without the JTD, the
system is insulating and the band gap is 0.3 eV. This value is however smaller than
the estimate $0.7$ eV expected from the tight-bonding analysis (the value of
effective two-center $dd\sigma$ integral). The difference is caused by the additional
broadening of the bonding and nonbonding bands in the LSDA. The JTD
increases the band-gap (for example, for $\delta$$=$$0.03$ the band-gap
is $0.7$ eV). Second, the partial densities of states in the bonding
part of the spectrum clearly shows the characteristic peaks at the band
edges, which is the signature of (quasi-) one-dimensional behavior.
The bandwidth is however larger than
$(\sqrt{3}$$-$$1)$$\times$$0.7$$\simeq$$0.5$ eV,
expected from the TB analysis. The density of nonbonding and antibonding
state is modified more significantly. Particularly, the nonbonding and antibonding
parts of the spectrum
overlap with the minority-spin $t_{2g}$ band (not shown in Fig. \ref{fig.LSDAdos}).
The minority-spin $t_{2g}$ states may interact with the majority-spin $e_g$ states
of the neighboring zigzag chains either directly, via next nearest-neighbor
hoppings (note, that the nearest-neighbor hoppings between the $t_{2g}$ and $e_g$
orbitals are forbidden in the cubic lattice -- Ref. \onlinecite{SlaterKoster}),
or indirectly, via the oxygen $2p$ band. As the result, the
''nonbonding''
states associated with the corner atoms $B$ and $B'$
are significantly broadened (to be compared with the $\delta$-peak
expected
from the TB analysis - Fig. \ref{fig.tbdos}).
Conversely, the $y^2$$-$$z^2$ and $x^2$$-$$z^2$
states of the bridge sites $A$ and $A'$ are only weakly bonded, resulting in the high
peak of the density of states around $0.5$ eV.

  Due to the hybridization, the distribution of the majority-spin
Mn($4p$) states around the Fermi level repeat the characteristic
bonding-nonbonding-antibonding splitting of the Mn($3de_g$) states.
The nonbonding part of the spectrum is composed mainly of the $x$ ($y$) orbitals
of the bridge atoms $A$ ($A'$).
The contribution of the $4p$ orbitals of the corner atoms into this region is
significantly smaller.

\begin{figure}
\centering \noindent
\resizebox{10cm}{!}{\includegraphics{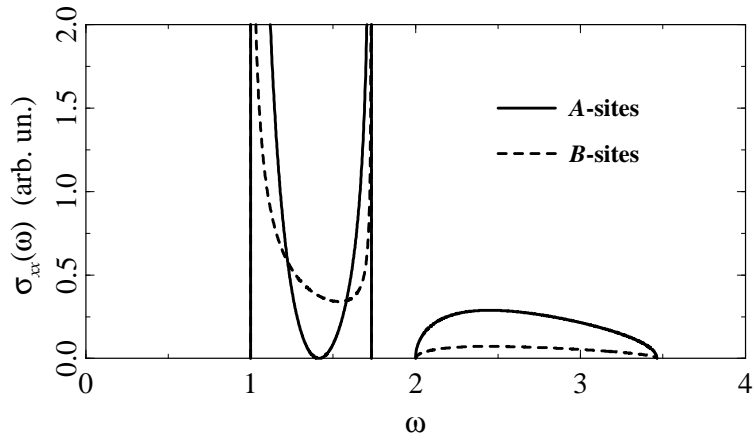}}
\caption{Partial contributions to the conductivity tensor of the
         tight-binding model.}
\label{fig.opt.tb}
\end{figure}

\begin{figure}
\centering \noindent
\resizebox{12cm}{!}{\includegraphics{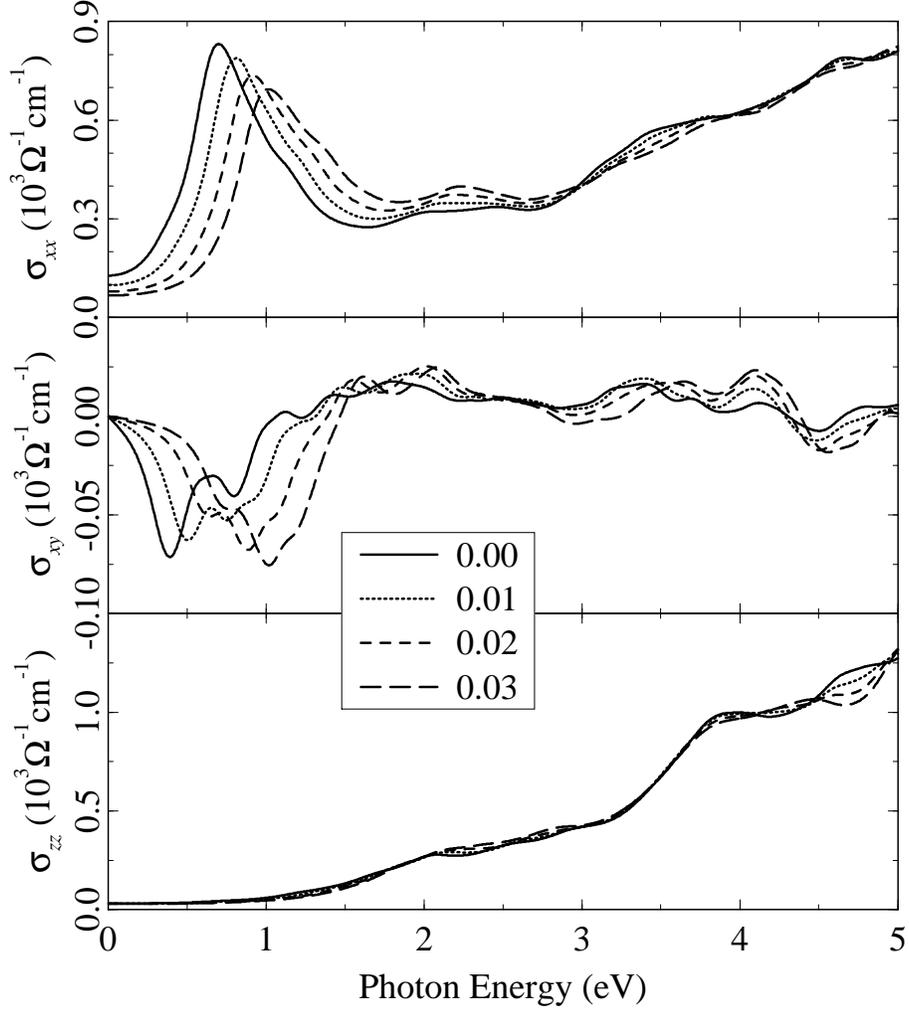}}
\caption{Elements of the conductivity tensor for Y$_{1/2}$Sr$_{3/2}$MnO$_4$
         in the zigzag antiferromagnetic state. Different lines correspond to
         different values of the Jahn-Teller distortion $\delta$.}
\label{fig.opt.LSDA}
\end{figure}

  The corresponding optical conductivity is shown in Fig. \ref{fig.opt.LSDA}.
Some details of calculations can be found in Refs. \onlinecite{review.2,Solovyev5}.
Here we only would like to mention that in order to speed up rather heavy calculations
(note that taking into consideration the zigzag AFM ordering,
the unit cell of Y$_{1/2}$Sr$_{3/2}$MnO$_4$ consists of 56 atoms), the Brillouin zone
integration in the calculations of optical conductivity was replaced by a summation
with the phenomenological Lorentzian broadening of $0.136$ eV. This causes some artificial
broadening of the spectrum. Nevertheless, we believe that all important features are
preserved in it, and were not washed out by the broadening. As the test, one can consider
the off-diagonal element $\sigma_{xy}(\omega)$ in the low-energy part of the spectrum.
It clearly shows two peak around $0.4$ eV and $0.8$ eV (the case of $\delta$$=$$0$),
corresponding to two peaks of the (quasi-) one-dimensional density of states in the
bonding region (Fig. \ref{fig.LSDAdos}), in a close analogy with the TB
picture (Fig. \ref{fig.opt.tb}).

  Surprisingly however that
the shape of diagonal conductivity in the $xy$-plane
is very different, contrary to our expectations based on the TB model.
In the low-energy part of the spectrum it has only one large peak at $0.7$ eV.
The difference cannot be simply attributed to the broadening effects,
because otherwise similar
broadening would be expected also for the off-diagonal component of the conductivity
tensor. Unfortunately,
on the level of first-principles band structure calculations, it
is rather difficult to decompose the optical conductivity unambiguously
into partial contributions and to elucidate the origin of the spectral shape.
Most probably, the difference between the LSDA calculations and the TB analysis
is caused at once by several factors outlined in the beginning of this section.

  In the TB model,
two contributions to the conductivity spectrum
in the region of bonding-nonbonding transitions,
$\sigma_{xx}^A(\omega)$ and
$\sigma_{xx}^B(\omega)$, are comparable (see Fig. \ref{fig.opt.tb}).
The off-diagonal element of the conductivity tensor, which is responsible
for the anisotropy of optical properties, is related
with the diagonal one by
Eq. (\ref{eqn:ocBsym.1}).
Therefore, $\sigma_{xy}^B(\omega)$ is expected to be of the same order of
magnitude as the total diagonal conductivity
$\sigma_{xx}^A(\omega)$$+$$\sigma_{xx}^B(\omega)$.
However, in a more realistic situation, when the system is not strictly one-dimensional,
the optical anisotropy is significantly reduced.
For example, in the LSDA we have
only
$|\sigma_{xy}(\omega)|$$\sim$$0.1\sigma_{xx}(\omega)$
in the region of bonding-nonbonding transitions.

  The JTD around the bridge sites results in the upward shift
of the low-energy peak and the simultaneous decrease of its intensity. Two experimental
groups reported somewhat different results for the peak position at the low temperature:
$1.0$ eV (Ref. \onlinecite{Jung}) and $1.3$ eV (Ref. \onlinecite{Ishikawa}).
Both are larger than $0.7$ eV obtained in the LSDA calculations without the
JTD. The JTD $\delta$$=$$0.03$
yields the new peak position $1.0$ eV, in agreement with one of the experimental
reports. However, the obtained intensity $700$ $\Omega^{-1}{\rm cm}^{-1}$ at
the peak maximum appears to be smaller than the experimental one, which is about
$1100$ $\Omega^{-1}{\rm cm}^{-1}$.\cite{Jung,Ishikawa}
Another effect of the
JTD is the redistribution of the in-plane
conductivity
around $2.0$ eV, in the region of bonding-antibonding transition.

  As expected for the layered perovskite compound,
the in-plane
($\sigma_{xx}$) and out-of-plane ($\sigma_{zz}$) components of the optical conductivity
reveal a strong anisotropy, in a good agreement with the experimental finding
for La$_{1/2}$Sr$_{3/2}$MnO$_4$.\cite{Jung}

  Finally, we calculate nearest-neighbor magnetic interactions in
Y$_{1/2}$Sr$_{3/2}$MnO$_4$
as the function of the Jahn-Teller distortion $\delta$, using the same approach
as in Ref. \onlinecite{Solovyev7}. For small $\delta$
we obtain $J_{A_1B_1}(\delta)$$\simeq$$43$$+$$313\delta$ meV and
$J_{A_1B_2}(\delta)$$\simeq$$12$$-$$680\delta$ meV.
Therefore, even without the JTD, the magnetic interactions in the
zigzag AFM state show a very strong anisotropy
$J_{A_1B_1}(0)$$>$$J_{A_1B_2}(0)$, in a good agreement with the
TB analysis. In the LSDA, a small JTD $\delta$$\geq$$0.02$ becomes
indispensable to stabilize the AFM coupling $J_{A_1B_2}(\delta)$$<$$0$
between neighboring zigzag chains.

\section{Summary}
\label{sec:summary}

  Properties of the so-called ''charge-ordered'' manganites are typically
regarded as one of the most striking examples of the strong coupling amongst
the spin, charge and orbital degrees of freedom. The coupling, which is reflected
in the number of transport, structural and magnetic phenomena, is also believed
to be responsible for the peculiar form of the ordered state realized at
low temperatures.

  In the present work we tried to elucidate the cause of such an
unusual behavior and to understand it from the viewpoint of the magnetic spin
ordering. We found that the zigzag form of the antiferromagnetic spin
ordering, combined  with the DE physics, already provides
a very consistent description for various low-temperature properties of the
''charge-ordered'' manganites. In the DE limit, the one-dimensional
FM zigzag chains become effectively separated from each other.
The one-dimensional character of the chains leads to the anisotropy of
electronic properties. The anisotropy of nearest-neighbor magnetic
interactions in the DE limit, combined with the isotropic
superexchange interaction $J^S$, readily explains the local stability of the
zigzag AFM state. Formally, the basic DE picture is
sufficient to understand the local stability. The JTD significantly
reduces the FM DE contribution to the AFM
bonds, and in this way widens the region of local stability of the
zigzag AFM ordering. It appears that in the wide range of parameters
$J^S$, the zigzag AFM ordering is stable not only locally, but
also globally, meaning that this is indeed the magnetic ground state of the system.
We also expect the anisotropy of optical properties, which
is related with the distribution of the $e_g$
states at the corner atoms of the zigzag chain.
The two-fold degeneracy of the $e_g$ levels plays a very important role in the
problem, and is responsible for the insulating behavior and the orbital ordering
developed
at the bridge sites of the zigzag chain.

  Thus, we believe that the nature of the low-temperature CE state
in manganites
is magnetic.
The charge ordering reduces mobility of the $e_g$ electrons and tends
to destroy the ferromagnetic coupling within the chain. Therefore, it is
expected to be small and play only minor role in the problem.

  Applicability of this picture to the high-temperature regime is still
in question. Particularly, many charge ordered manganites are characterized by
the existence of two transition temperatures: the long-range AFM ordering occurs
below the N\'{e}el temperature $T_{\rm N}$, whereas the charge and orbital ordering
orderings occur simultaneously below another critical temperature
$T_{\rm CO}$. In accordance with neutron scattering measurements, it holds
$T_{\rm N}$$<$$T_{\rm CO}$ (see, e.g., Ref. \onlinecite{Sternlieb}). This fact
is typically considered as the strong experimental evidence
that the charge and orbital
degrees of freedom play a decisive role in the interval
$T_{\rm N}$$<$$T$$<$$T_{\rm CO}$ and drive the AFM CE ordering at
$T_{\rm N}$.\cite{Zimmermann,comment.9}
However,
other experimental data suggest that
the exact nature of the state realized between $T_{\rm N}$ and
$T_{\rm CO}$ is much more complicated, and still far from the complete
understanding.\\
1) Some magnetic experiments simply ''do no see'' the existence of the
transition point $T_{\rm N}$. For example, the behavior of spin magnetization
in the external magnetic field shows the characteristic metamagnetic transition
both below and above $T_{\rm N}$, which disappears only at
$T_{\rm CO}$.\cite{Tokunaga,Millange}
The magnetic susceptibility shows a pronounced peak at $T_{\rm CO}$ and no
anomaly at $T_{\rm N}$.\cite{Millange} These data suggest that, although
there is no long-range magnetic order, the magnetic fluctuations and the
effects of short-range order continue to play a very important role
in the interval $T_{\rm N}$$<$$T$$<$$T_{\rm CO}$.\\
2) The high-resolution electron microscopy clearly reveals the existence of the
one-dimensional chains (stripes) even above $T_{\rm N}$, though the
exact from of the stripes is not quite clear yet.\cite{Mori}
It is also not clear whether these stripes have a magnetic origin
or not. If they do, it would be a strong argument in the favor of the magnetic scenario
considered in the present work, although the perfect geometry of the FM
zigzag chain may be destroyed at elevated temperatures.
In this respect, an interesting theoretical
result has been obtained by Hotta {\it et~al}.
They argued that the class of one-dimensional zigzag objects, which preserves
the insulating behavior at $x$$=$$0.5$ is actually much wider and is not limited
by the zigzag chain shown in Fig. \ref{fig.chains}. All these new zigzag
configuration have higher energies, and therefore are not realized as the ground state
at low temperatures. However, they may carry a considerable weight
in the thermodynamics averages and therefore contribute to the high-temperature
behavior of the charge ordered manganites.

\section*{Acknowledgments}

I thank K. Terakura for valuable discussions.
The present work is partly supported
by New Energy and Industrial Technology
Development Organization (NEDO).

\appendix

\section{Local densities of states for the zigzag chain}
\label{subsec:dos}

  The local density of states at the site $\tau$ is given by
\begin{equation}
N^L_\tau(\varepsilon) = -\frac{1}{\pi} {\rm Im} G^{LL}_{\tau \tau} (\varepsilon).
\label{eqn:dos}
\end{equation}
Using the explicit expressions for the matrix elements of the Green function,
Eqs. (\ref{eqn:g11AA})-(\ref{eqn:g22BB}), noting that
$-$$\frac{1}{\pi} \lim_{\eta \rightarrow 0+} {\rm Im} [\varepsilon$$-$$\varepsilon'
$$+$$i\eta]^{-1}$$=$$\delta(\varepsilon$$-$$\varepsilon')$, and taking into account that
\begin{equation}
\int dk \delta(\varepsilon-\varepsilon_k^\pm)=\frac{1}{|d\varepsilon_k^\pm/dk|_\varepsilon},
\label{eqn:delta_property}
\end{equation}
where the derivative is evaluated at the point
$\varepsilon_k^\pm$$=$$\varepsilon$, one can find the following expressions
for the partial densities of states:
\begin{equation}
N^1_A(\varepsilon)=\frac{1}{\pi} \frac{|\varepsilon-\Delta_{\rm C}|}
{\sqrt{[3+\varepsilon(\Delta_{\rm C}-\varepsilon)][\varepsilon(\varepsilon-\Delta_{\rm C})-1]}},
\label{eqn:n11A}
\end{equation}
\begin{equation}
N^1_B(\varepsilon)=\frac{1}{2\pi|\varepsilon-\Delta_{\rm C}|} \sqrt{ \frac
{3+\varepsilon(\Delta_{\rm C}-\varepsilon)}{\varepsilon(\varepsilon-\Delta_{\rm C})-1}}
+ \frac{3-\sqrt{3}}{2} \delta(\varepsilon-\Delta_{\rm C}),
\label{eqn:n11B}
\end{equation}
\begin{equation}
N^2_B(\varepsilon)=\frac{3}{2\pi|\varepsilon-\Delta_{\rm C}|} \sqrt{ \frac
{\varepsilon(\varepsilon-\Delta_{\rm C})-1}{3+\varepsilon(\Delta_{\rm C}-\varepsilon)}}
+ \frac{\sqrt{3}-1}{2} \delta(\varepsilon-\Delta_{\rm C}).
\label{eqn:n22B}
\end{equation}
The total density of states, $N(\varepsilon)$$=$$N^1_A(\varepsilon)$$+$$
N^1_B(\varepsilon)$$+$$N^2_B(\varepsilon)$, is given by
\begin{equation}
N(\varepsilon)=\frac{1}{\pi} \frac{|2 \varepsilon-\Delta_{\rm C}|}
{\sqrt{[3+\varepsilon(\Delta_{\rm C}-\varepsilon)][\varepsilon(\varepsilon-\Delta_{\rm C})-1]}}
+ \delta(\varepsilon-\Delta_{\rm C}).
\label{eqn:ntotal}
\end{equation}

\section{The total energy change due to small
         rotations of the spin magnetic moments}
\label{apndx}

  Here we argue that the invariance of the total energy with respect to
uniform rotations of all spin magnetic moments by the same angle automatically
leads to the isotropic Heisenberg form for the total energy change
associated with small rotations of an arbitrary type -- Eq. (\ref{eqn:heisenberg}).
For simplicity we consider the FM ordering for which
${\bf e}^0_{\it i}$$\equiv$${\bf e}^0$ does not depend on the site ${\it i}$.
By expanding
$E_t[\{ {\bf e}^0_{\it i} + \mbox{\boldmath$\delta$}{\bf e}_{\it i} \}]$
in the Taylor series we obtain in the second order:
\begin{equation}
E_t[\{ {\bf e}^0 + \mbox{\boldmath$\delta$}{\bf e}_{\it i} \}] \simeq
E_t[ {\bf e}^0 ] - \sum_{\it i} f_{\it i}
(\mbox{\boldmath$\delta$}^{(2)}{\bf e}_{\it i} \cdot {\bf e}^0 )
-\frac{1}{2} \sum_{\it ij} J_{\it ij}
(\mbox{\boldmath$\delta$}^{(1)}{\bf e}_{\it i} \cdot
\mbox{\boldmath$\delta$}^{(1)}{\bf e}_{\it j} ).
\label{eqn:ap1}
\end{equation}
In Eq. (\ref{eqn:ap1}) we employed the requirement of rotational invariance
for the perturbed system: if all spins are rotated
the same angle,
the total energy does not change. Therefore, ${\bf e}^0$ and
$\mbox{\boldmath$\delta$}{\bf e}_{\it i}$ may enter the expression for the total
energy change only through the scalar products. Noting that
$\mbox{\boldmath$\delta$}^{(1)}{\bf e}_{\it i}$ and
$\mbox{\boldmath$\delta$}^{(2)}{\bf e}_{\it i}$
are given by Eq. (\ref{eqn:rotations})
and employing again the requirement of rotational invariance for
the unperturbed system (that is the total energy does not change if all
spins $\{ {\bf e}^0_{\it i} \}$  are rotated by the same angle
$\mbox{\boldmath$\delta\varphi$}_{\it i}$$\equiv$$\mbox{\boldmath$\delta\varphi$}$)
we find the following condition:
$$
f_{\it i} = \sum_{\it j} J_{\it ij}.
$$
This proves validity of Eq. (\ref{eqn:heisenberg}) for the FM ordering.
The generalization to the AFM ordering is
straightforward, because similar arguments can be applied both for the intra-
and inter-spin-sublattice interactions in the AFM state.


\begin{figure}
\centering \noindent
\resizebox{12cm}{!}{\includegraphics{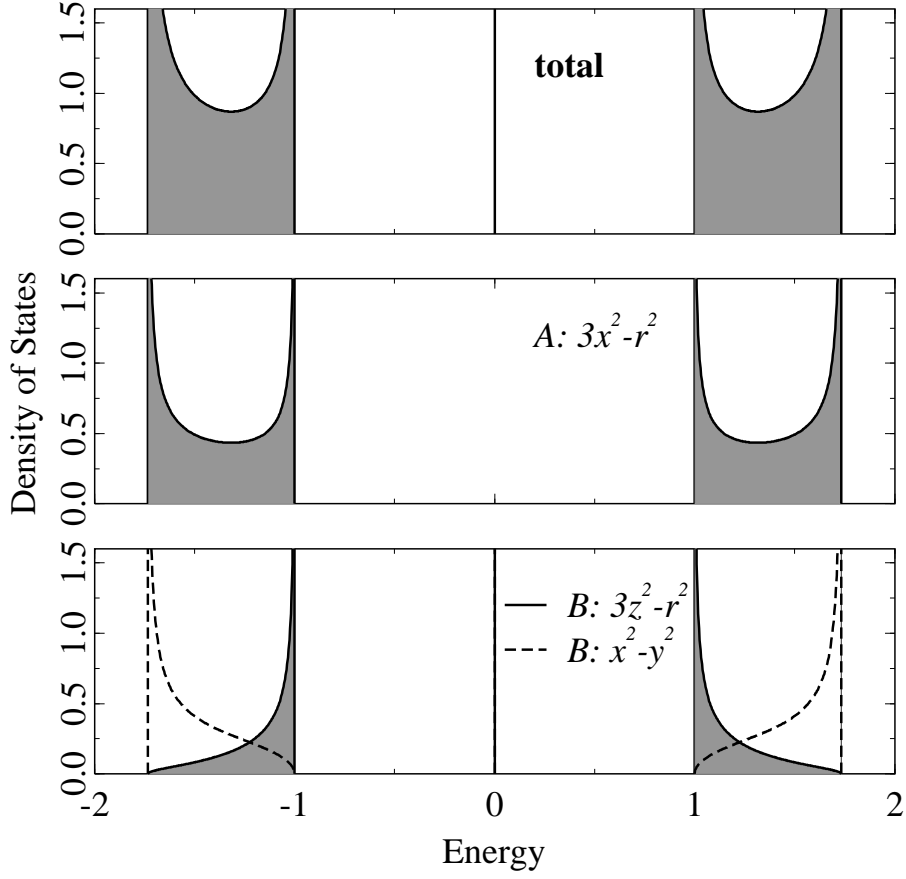}}
\caption{Total and partial densities of states of the tight-binding model
         for $\Delta_{\rm C}$$=$$0$ (no charge disproportionation).}
\label{fig.tbdos}
\end{figure}


\begin{thebibliography}{10}
\bibitem{comment.7}
Note that in the ''charge ordering'' regime $n_B$$=$$1$$-$$n_A$,
$n_A$$=$$-$$\frac{1}{\pi} \int_{-\infty}^{\varepsilon_F}
d\varepsilon G^{11}_{AA}(\varepsilon)$, and $G^{11}_{AA}(\varepsilon)$
is given by Eq. (\ref{eqn:g11AA}).
\bibitem{comment.8}
Since the pole of $G^{22}_{A_1A_1}(\varepsilon)$ is located in the unoccupied part of
the spectrum, the term proportional to $\left(G^{22}_{A_1A_1}(\varepsilon)\right)^2$ does not
contribute to the exchange coupling (\ref{eqn:JA1A3}) and can be dropped.
\bibitem{comment.1}
Here we consider only the ''hard'' condition of the local stability of the A-type
AFM ordering and require the sign of the nearest-neighbor
magnetic coupling
to correspond directly the type of the magnetic ordering.
For example, due to existance of
the longer-range interactions between the antiferromagnetically
coupled layers, the stability of the A-type AFM ordering with
respect to the spin canting is given by a ''softer'' condition, where
the lower boundary for $|J^S|$ is only $0.063$ when $x$$=$$0.5$.\cite{Solovyev3}
\bibitem{comment.6}
The canted spin states of the FA type have been considered in our previous
work (Ref. \onlinecite{Solovyev3}). The canted spin states of the AG type
can be generated along the same line and corresponds to the combination of angles
$(\theta_{xy},\pi)$, in the notations of Ref. \onlinecite{Solovyev3}.
If stable, the AG states have the lowest energy amongst the collinear FM,
A-, C-, and G-type AFM states and all possible intermediate canted states for
$x$$=$$0.5$.
\bibitem{comment.4}
We do not consider the on-site interactions between the $3de_g$ and $4p$ orbitals,
which are caused by the nonsphericity of the Coulomb potential. The effect is
believed to be small.\cite{Elfimov}
In this respect, results of the LSDA band structure calculations presented in
Sec. \ref{sec:LSDA} correspond to the same level of approximation: the
calculations were performed in the ASA geometry,\cite{Andersen}
meaning that at each atomic site the Coulomb and exchange potentials were
averaged over the angles within the atomic sphere.
\bibitem{comment.9}
Though, this is not the only one scenario. Another possibility is the
temperature transition of the CE-type AFM state to the homogeneous FM
state, which is observed for example in
Pr$_{0.5}$Sr$_{0.5}$MnO$_3$ and Nd$_{0.5}$Sr$_{0.5}$MnO$_3$.\cite{review.1}
This schenario seems to be consistent with the magnetic origine of the
CE state.
\end{thebibliography}
\end{document}